\documentclass[reprint,aps,prl,showpacs,superscriptaddress]{revtex4-1}
\usepackage{amsmath}
\usepackage{amssymb}
\usepackage{graphicx}
\usepackage{subcaption}
\usepackage{float}
\usepackage{bm}
\usepackage{indentfirst}
\usepackage[colorlinks=true,linkcolor=blue,citecolor=blue,urlcolor=blue]{hyperref}
\usepackage[usenames,dvipsnames]{xcolor}
\usepackage{siunitx}
\usepackage[normalem]{ulem}
\usepackage{color}
\begin{document}
\title{The Quantum Anomalous Hall Majorana Platform}
\author{Yongxin Zeng}
%\affiliation{ICQD, Hefei National Laboratory for Physical Sciences at Microscale, and Synergetic Innovation Center of Quantum Information and Quantum Physics, University of Science and Technology of China, Hefei, Anhui 230026, China.}
\affiliation{CAS Key Laboratory of Strongly-Coupled Quantum Matter Physics, and Department of Physics,  University of Science and Technology of China, Hefei, Anhui 230026, China.}
\author{Chao Lei}
\affiliation{Department of Physics, The University of Texas at Austin, Austin, Texas 78712, USA}
\author{Gaurav Chaudhary}
\affiliation{Department of Physics, The University of Texas at Austin, Austin, Texas 78712, USA}
\author{Allan H. MacDonald}
\affiliation{Department of Physics, The University of Texas at Austin, Austin, Texas 78712, USA}
\date{\today}
\begin{abstract}
We show that quasi-one-dimensional (1D)
quantum wires can be written onto the surface of magnetic topological insulator (MTI)
thin films by gate arrays.
When the MTI is in a quantum anomalous Hall (QAH) state, MTI$/$superconductor (SC) quantum wires have
especially broad stability regions for both topological and non-topological states,
facilitating creation and manipulation of Majorana particles on the MTI surface.
\end{abstract}
\maketitle
%\section{Introduction}

\noindent
\textit{Introduction:}
Non-Abelian anyons that can be localized at controllable
positions provide an attractive platform for fault-tolerant quantum computation \cite{Ivanov2001,Nayak2008,Kitaev2003,Kitaev2006,Sarma2015,Aasen2016,Karzig2017}.
Candidate non-Abelian particles that have already been realized in solid
state systems include fractionally charged excitations of the $\nu = 5/2$ fractional quantum Hall
liquid \cite{Moore1991}, Abrikosov vortices in two-dimensional topological
superconductors \cite{Sun2017}, and Majorana zero mode (MZM) end states of
$p$-wave superconducting quantum wires \cite{Kitaev2001,Jason2012,Lutchyn2010,Sau2010,Sau2010}.
In this article we propose a new strategy for creation of MZMs
and manipulations of MZM positions
on a two-dimensional surface.

\begin{figure}
	\centering
	\begin{subfigure}{\linewidth}
		\includegraphics[width=0.9\linewidth]{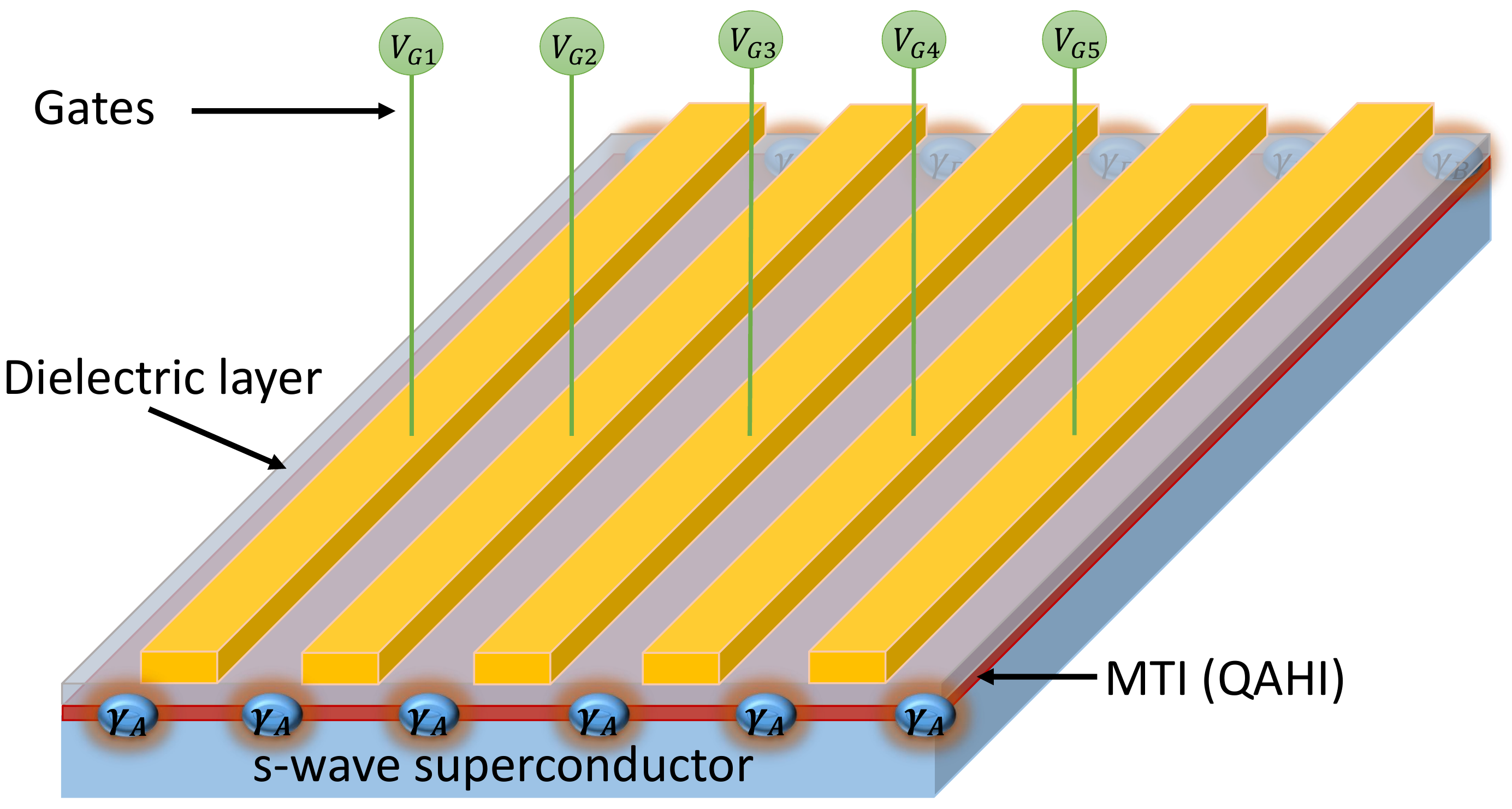}
		\subcaption{} \label{MajoranaSetup}
	\end{subfigure}
	\begin{subfigure}{\linewidth}
		\includegraphics[width=\linewidth]{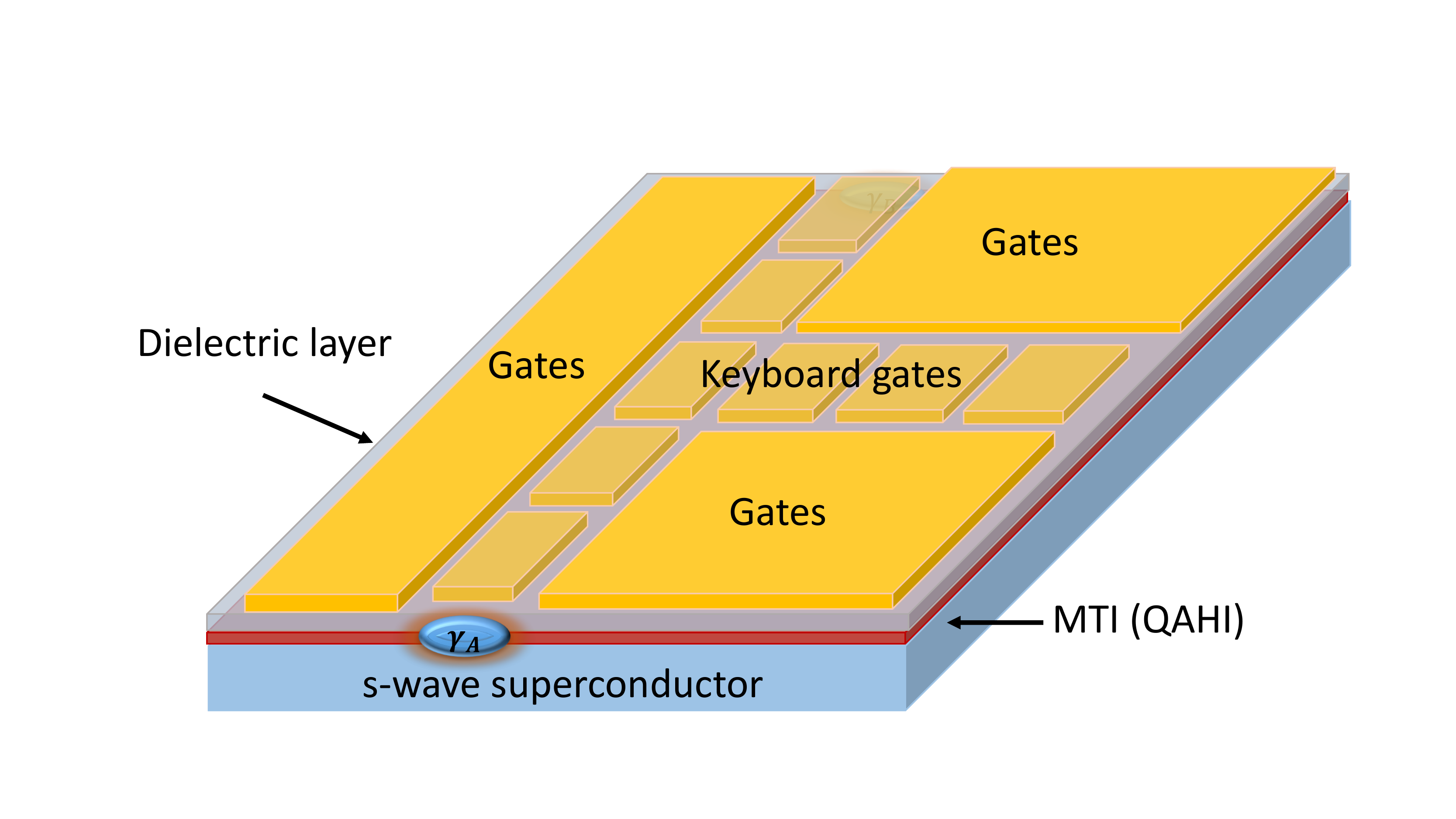}
		\subcaption{} \label{Tsetup}
	\end{subfigure}
	\caption{ (a) Majorana zero modes can be created locally by separating the surface of a magnetic topological insulator in a Quantum Anomalous Hall state into alternating normal and topological regions using remote gates;
		(b) A T-junction can be defined and controlled by local gates to achieve manipulation and braiding of Majoranas.}
		\label{setup}
\end{figure}

MZMs appear at boundaries between normal and topological
quasi-1D superconductors \cite{Kitaev2001,Lutchyn2010,Sau2010,Beenakker2013}.
The simplest 1D topological superconductor (TSC)
consists of spinless electrons with near-neighbor attractive interactions \cite{Kitaev2001}.
Topological superconducting states can occur, however, in any quasi-1D system with
superconductivity and broken time-reversal symmetry \cite{Jason2012}.
Much progress has been achieved both theoretically \cite{Fu2008,Lutchyn2010,Sau2010}
and experimentally \cite{Mourik2012,Deng2012,Das2012,Deng2016,Rokhinson2012,Albrecht2016}
by studying strongly spin-orbit-coupled proximitized quantum wires with large $g$-factors
that are perturbed by an external magnetic field, and also
by placing magnetic-atom arrays on superconducting substrates \cite{Nadj2014,Pawlak2016}.
Here we suggest an alternate approach, based on the QAH state \cite{Yu2010,Chang2013,He2017}
of MTIs, that has potentially valuable advantages. We show that ribbons formed from
MTIs are often topological when proximity-coupled to a superconductor \cite{Law2017}.
MZMs also appear
when quasi-1D topological regions are written onto a portion of the surface of an MTI by external gates,
as illustrated schematically in Fig.~\ref{setup}.
Furthermore, the MTI$/$SC phase diagram has particularly broad
regions of stability for both normal and topological phases when the isolated MTI is
close to a phase boundary between normal insulator and QAH states \cite{Yu2010,Chang2013}.
This property provides a practical route towards electrical control of Majorana positions, and braiding operations via Majorana T-junctions \cite{Alicea2011}, and recently proposed Majorana box qubit \cite{Plugge2017, Karzig2017}.

\noindent
\textit{Theoretical model of a proximized MTI:}
We focus on Cr-doped Bi$_2$Se$_3$ thin films
in which proximitized superconductivity has already been
demonstrated \cite{Chang2013,He2017} experimentally,
and which are close to the MTI's QAH insulator/normal insulator phase boundary when 4-6
quintile layers thick.
In the
$ \phi_{\bm k} = (c_{\bm{k}\uparrow}^t,c_{\bm{k}\downarrow}^t,c_{\bm{k}\uparrow}^b,c_{\bm{k}\downarrow}^b)^T$
basis,
where $ c_{\bm{k} \sigma}^s$ annihilates an electron with
momentum $ \bm{k} $, spin $ \sigma = \uparrow, \downarrow $ and surface $s= t/b $ (top/bottom),
the single particle Hamiltonian of a MTI thin film at energies below the bulk gap is
\begin{equation} \label{H0(k)}
H_0(\bm k) = h_{\textrm{D}}(\bm k)\tau_z + m_k\tau_x + \lambda\sigma_z + \lambda'\tau_z
\end{equation}
where $\tau_{i}$ acts on surface and $\sigma_{i}$ on spin.
In Eq.~\ref{H0(k)} $h_{\textrm{D}}(\bm k)=v(k_y\sigma_x-k_x\sigma_y)$ is the 2D Dirac isolated
surface Hamiltonian with Fermi velocity $v$;
$m_k=m_0+m_1(k_x^2+k_y^2)$ accounts for hybridization between
top and bottom surfaces, $\lambda$ is the exchange field produced by the ferromagnetically ordered
magnetic dopants; and $\lambda'$ accounts for the energetic displacement between
Dirac cones on top and bottom surfaces produced by vertical electric fields in the bulk of the TI.
When placed in proximity to an $s$-wave SC, the system is
described by the Bogoliubov-de Gennes (BdG) Hamiltonian
$H(\bm k) = \sum_{\bm k} \psi_{\bm{k}}^\dagger H_{\textrm{BdG}}(\bm k)\psi_{\bm{k}}/2$, where

\begin{equation}
\label{HBdG}
H_{\textrm{BdG}}(\bm k) = \left(\begin{array}{cc}
H_0(\bm k)-\mu & \Delta_{\textrm{sc}} \\
\Delta_{\textrm{sc}}^\dagger & -H_0^*(-\bm k)+\mu
\end{array}\right)\\
\end{equation}
%\end{equation}
with
\begin{equation}
\Delta_{\textrm{sc}} = \left(\begin{array}{cc}
i\Delta_t\sigma_y & 0 \\
0 & i\Delta_b\sigma_y
\end{array}\right).
\end{equation}
and
\begin{equation}
\psi_{\bm{k}} = (c_{\bm{k}\uparrow}^t,c_{\bm{k}\downarrow}^t,c_{\bm{k}\uparrow}^b,c_{\bm{k}\downarrow}^b,c_{-\bm{k}\uparrow}^{t\dagger},c_{-\bm{k}\downarrow}^{t\dagger},c_{-\bm{k}\uparrow}^{b\dagger},c_{-\bm{k}\downarrow}^{b\dagger})^T
\end{equation}
A regularized square-lattice version of Eq.~\ref{HBdG} with nearest neighbor
hopping \cite{Marchand2012} accurately and conveniently captures
the physics of topological phase transitions (TPT) at the surface of MTI.
In the lattice model $k_{x/y}$ is replaced by $\sin k_{x/y}$ and
$k^2$ by $2(2-\cos k_x - \cos k_y)$.
We absorb the lattice constant
into the wavevectors in Eq.~\ref{H0(k)} to
make them dimensionless, so that $m_0,m_1$ and $v$ all have dimensions of energy.
The inter-surface hybridization parameter $m_1$ plays an essential role
by preventing the appearance of unphysical states at low energies away from $\bm{k}=0$.
For $m_0 \ll m_1$ only the $\Gamma$-point avoided crossing
is relevant at low energies.% one, and the other high symmetry points are at much higher  energies.
We can obtain realistic values of $v$, $m_0$ and $m_1$ by comparing the model spectrum
with DFT band structures of  Bi$_2$Se$_3$, like the one illustrated Fig. S1 of
the Supplemental Material \cite{supplement}.
For a five-layer film we find that the gap at
$\Gamma$ is about $\SI{12}{\milli\electronvolt}$, giving $m_0=\SI{6}{\milli\electronvolt}$.
Rough estimates for $m_1$ and $v$ can be obtained by fitting to the DFT Dirac velocity $v$
($ \sim 4 \times 10^{5} \si{m/s}$) and the DFT gaps at $M$ and $K$. We find that
$m_1 \sim \SI{0.2}{\electronvolt}$ and $v \sim \SI{0.7}{\electronvolt}$ (notice that the lower band of DFT result gives a different value of $v$. As shown in the Supplemental Material \cite{supplement}, our main findings on topological robustness are independent of upper or lower band Fermi velocity as long as $v$ and $m_1$ remains on larger energy scale compared to $m_0$) for a lattice model with
lattice constant $a \approx 4$\AA.

\noindent
\textit{Topological classification of 2D bulk and 1D ribbon states:}
The 2D classification of MTI$/$SC states presented in Ref.~\onlinecite{Wang2015}
demonstrates the possibility of BdG bands with odd Chern numbers.
We first address the special case $\Delta_t=-\Delta_b=\Delta$
and set $\lambda'=\mu=0$ to obtain simple analytical expressions
for phase boundaries.
When the basis transformation used in Ref.~\onlinecite{Wang2015} is employed
the $8\times 8$ BdG Hamiltonian block diagonalizes into four $2\times2$ matrices,
each of which has the form of a spinless $p\pm ip$ superconductor.
Gaps close at $\Gamma$ for $\pm\lambda\pm\Delta=m_0$.
For $|\lambda| > m_0$, {\it i.e.} when the QAH
effect occurs, weak pairing gives rise to superconducting states with Chern numbers $\pm 2$,
corresponding to Nambu-doubled quantum Hall edge states.
As explained below, we propose using these states as a resource for MZM formation.
(The Chern number phase diagram of our proximitized MTI model is
described in detail in the Supplemental Material \cite{supplement}.)

\begin{figure}
	\centering
	\includegraphics[width=0.9\linewidth]{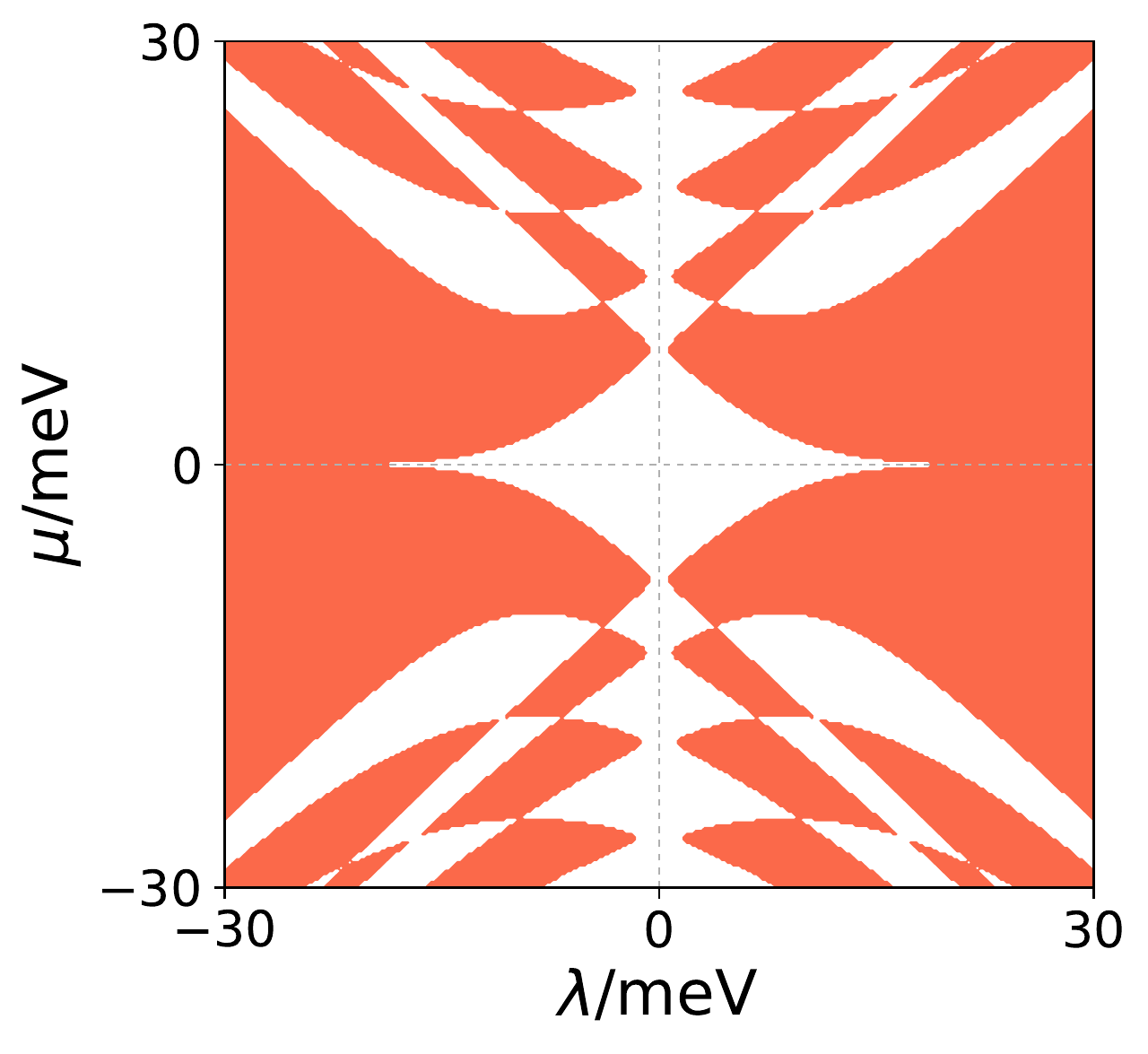}
	\caption{$Z_2$ ribbon phase diagram in $(\lambda,\mu)$ space for
		$\lambda'=0$, $m_0 =6$ meV, $m_1 = 200$ meV, $\Delta_{b}=1$ meV,
		$\Delta_{t}=0$, ribbon width $W=300$,
		and velocity $v = 700$ meV.  The orange regions are topologically non-trivial and
		support MZMs.}
	\label{lambdamu}
\end{figure}

%\subsection{Quasi-1D Chain}
Our main interest is in the possibility of gate-defined quasi-1D topological regions
embedded within the 2D film surface.  To set the stage, however, we first address the
closely related case of narrow ribbons \cite{Potter2010,Kitaev2001,Tewari2012}.
For width $W=1$, where $W$ is the number of lattice
model rows, the $\Delta_t=-\Delta_b$ BdG Hamiltonian
reduces to four effective Kitaev chains.
% with hopping terms that have different functional
%dependences on $m_0, m_1, \lambda$, $\Delta$ and different chemical potentials.
%Each yields MZMs localized near the chain ends when topologically non-trivial.
%However, a close by Majorana coming from the non-trivial Kitaev chain of a different block can hybridize to give a trivial electronic mode.
%A Kitaev chain is topologically non-trivial when $\nu$ is  occurs when the absolute value of chemical potential is equal to the nearest neighbor hopping amplitude.
%The four effective Kitaev chains have different effective chemical potential, and hopping terms with different functional dependence on $m_0, m_1, \lambda$, and $\Delta$ \ref{App:BlockDiag},
%which can facilitate a scenario with odd number of chains in the non-trivial phase with one pair of isolated MZM.
The Supplemental Material \cite{supplement} contains a detailed summary of how
the $Z_2$ (1D) topological phase diagram of the $\Delta_t=-\Delta_b$ model
depends on $W$. In Fig. \ref{lambdamu} we plot the $Z_2$ phase diagram in $(\lambda,\mu)$
space for a ribbon with proximity coupling only to the bottom MTI surface.
The $Z_2$ invariant is evaluated using
$ \nu = \textrm{sgn}[\textrm{Pf}\tilde{H}(0)] \textrm{sgn}[\textrm{Pf}\tilde{H}(\pi)]$,
where $ \tilde{H} $ is the skew-symmetrized Hamiltonian obtained by switching $H_{BdG}$ to the Majorana basis,
Pf denotes the Pfaffian number and $\textrm{sgn}[x]$ is the sign of $x$. More information about the calculation of $Z_2$ invariant is shown in the Supplemental Material \cite{supplement}.
For weak pairing $\nu=\pm1$ depending on whether the number of normal state bands that
cross the Fermi level is even or odd.

%When $W$ is large enough, the outline of the diagram looks very similar to that in the 2D case (Fig. \ref{N_lam_Del_lattice}).
%The $Z_2$ invariant becomes more and more sensitive to all parameters
%with increasing ribbon width when the chemical potential is inside the 2D bands.
%There are in $4W$ phase boundaries at which the quasi-1D $Z_2$-classification
%changes. (See Fig. \ref{N_lam_Del_lattice}).

% d number of MZM pairs at the chain ends, and will remain
%non-trivial for sufficiently weak coupling between the effective chains.

%In real solid state systems, ideal 1D chains are very difficult to form.
%A more practical way to realize such systems is to form quasi-1D chains.
%Multi-channel quasi-1D $p\pm ip$ superconductors can \cite{Potter2010,Kitaev2001} also
%host MZMs.
%Here we discuss the connection between TPTs with respect to dimensional reduction and corresponding Majorana edge or end mode.
%With a view toward explaining our strategy for writing robust MZMs into the surface of a MTI,
%we first recall\cite{Tewari2012} the 1D topological properties of finite width 2D $p\pm ip$ 2D superconductors.
%% The topic has been discussed in context of topological invariants\cite{Tewari2012}, while
%We focus on the evolution of the gap spectrum and on the relationship between 2D edge states
%and 1D MZMs.
%The system has three length scales, the length $L$, the width $W$, and the 2D edge
%state localization length $\xi_P$.  The 2D system is in class $D$, characterized by a
%topological invariant which counts the number of chiral Majorana states.

The broad region of topologically nontrivial behavior at large $\lambda$ and small $\mu$ in Fig.~\ref{lambdamu},
{\it i.e.} when the unproximitized MTI is in a QAH state, reflects the property that only a single
ribbon band is present in this energy range at any value of $W$.
The $Z_2$ invariant is qualitatively more sensitive to $\lambda$, $\mu$, $W$ and other
ribbon model parameters at larger values of $\mu$ that lie within the gapped surface state bands.
For a lattice model of a normal 2D p-wave superconductor, for example,
there are $4W$ phase boundaries in the quasi-1D $Z_2$ classification of ribbon states at width $W$.
This sensitivity is not favorable for reliable realization of either trivial or
non-trivial states. In the QAH state there is at most a single band, but
the MZMs present for finite ribbon length
are protected only by exponentially small superconducting gaps ($\sim \Delta e^{-W/\xi}$),
where $\xi \sim v/(\lambda - m_0)$ is the 2D edge state localization length in lattice constant units,
making the $Z_2$ classification academic.  (For typical parameters $\xi \sim 10 \si{nm}$)
In order to obtain MZMs that are reasonably localized near ribbon edges it is necessary to
have ribbon widths that are not too large compared to $\xi$, and also to be able to conveniently tune
between topologically trivial and non-trivial states.  Since the exchange coupling parameter
$\lambda$ in Fig.~\ref{lambdamu} is fixed for a given MTI sample and a given operating
temperature, a different tuning parameter must be identified.

%\item Next Section:  Displacement Field, carrier Density phase diagram - can move along lines
%in this diagram with a back gate
\noindent
\textit{Quantum Anomalous Hall Ribbon Majoranas:}
We propose controlling MZMs in ribbons by placing a MTI film that supports a QAH state
on a superconducting substrate and fabricating a top gate. Varying the gate field will alter
the carrier density (and hence the chemical potential $\mu$) of the MTI, and also shift
the energy of the top surface Dirac cone relative to the bottom surface.
The latter effect is due to the unscreened portion of the gate electric field
that survives in the interior of the TI and is represented in our model by the
parameter $\lambda'$.  Applying a gate voltage moves the system along a line
in $(\mu,\lambda')$ space that depends on the effective bulk dielectric constant of the TI.
As we now show a gate voltage can therefore tune the proximitized MTI between
$Z_2=0$ and $Z_2=1$ ribbon states.

\begin{figure*}
\centering
\begin{subfigure}[b]{0.3\textwidth}
\includegraphics[width=\textwidth]{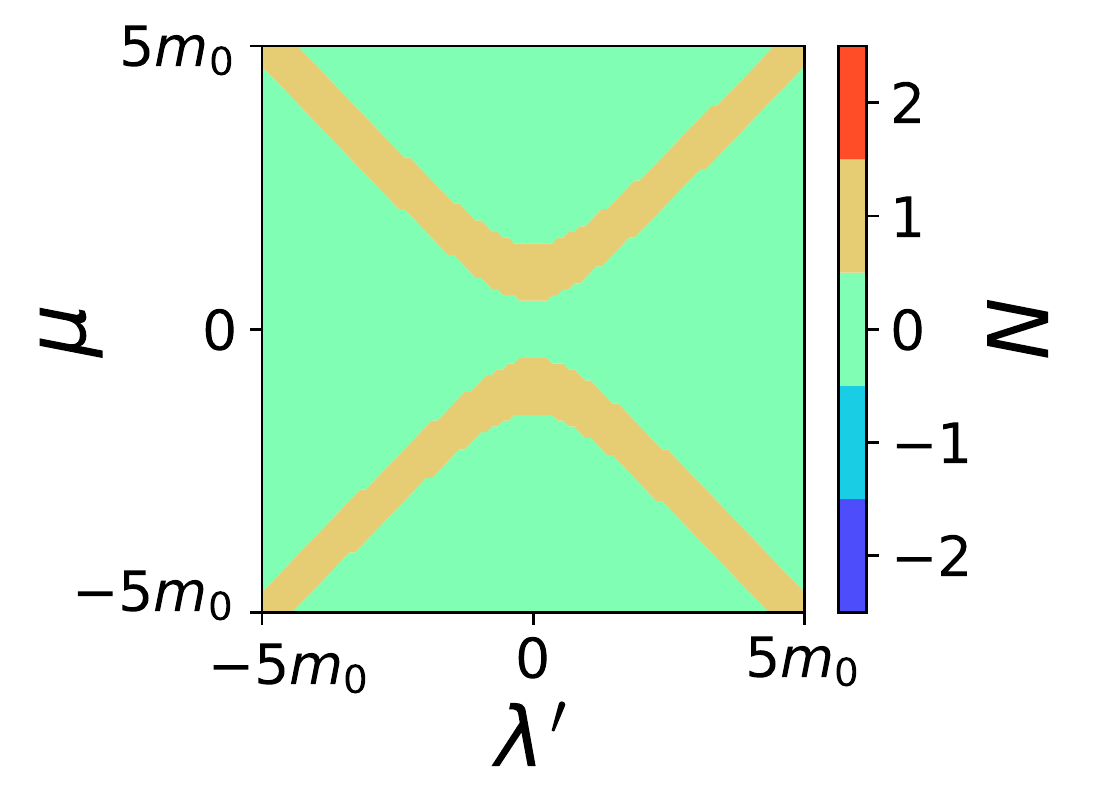}
%\subcaption{} \label{3a}
\end{subfigure}
\begin{subfigure}[b]{0.3\textwidth}
\includegraphics[width=\textwidth]{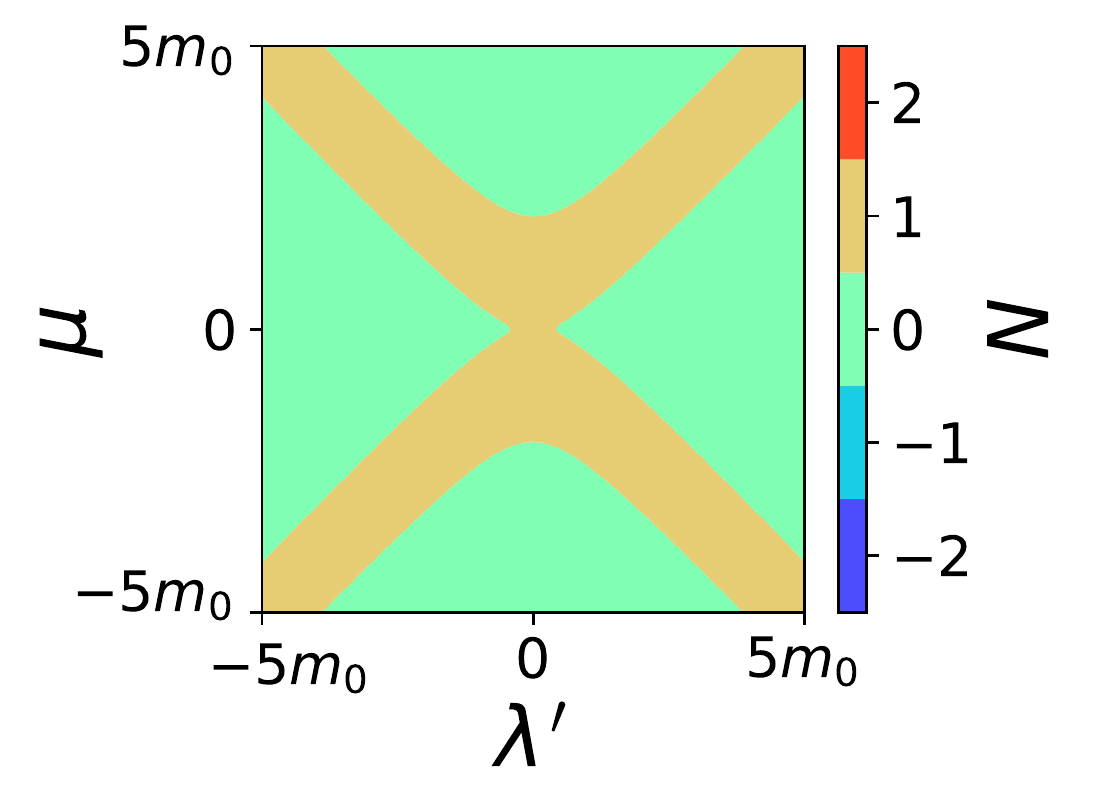}
%\subcaption{} \label{3b}
\end{subfigure}
\begin{subfigure}[b]{0.3\textwidth}
\includegraphics[width=\textwidth]{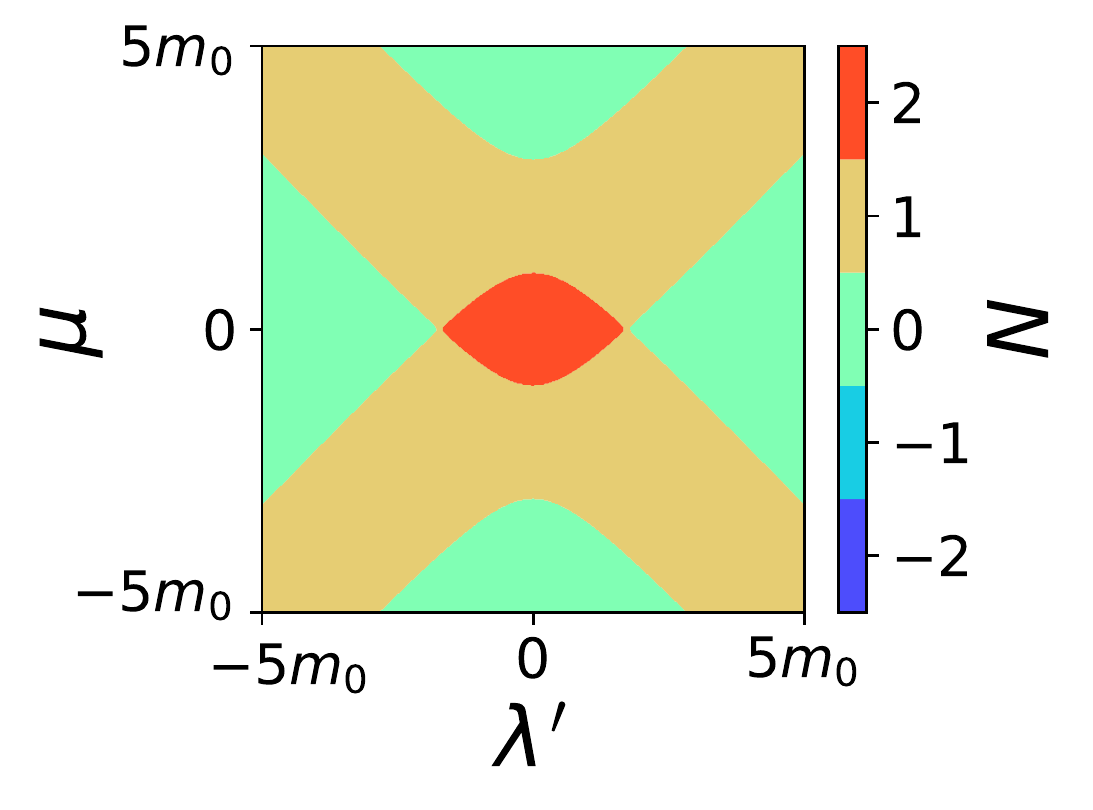}
%\subcaption{} \label{3c}
\end{subfigure}
\caption{Phase diagram of a MTI on a superconducting substrate in $\lambda'$-$\mu$ space.
Here $\Delta_b=0.2m_0$, $\Delta_t=0$, and from left to right
$\lambda= 0.5m_0$, $m_0$, and $2m_0$.}\label{phase_general}
\end{figure*}

Fig. \ref{phase_general} illustrates 2D Chern number phase diagrams in ($\lambda'$,$\mu$) space
for MTIs on superconducting substrates with $\Delta_b=0.2m_0$ and $\Delta_t=0$,
at three different values of exchange field $ \lambda $.
For $\lambda < m_0$ (left panel), the unproximitzed MTI is in a normal insulator state, but superconductivity
that is sufficiently strong can still induce odd Chern number
BdG states. Our main interest is in the case in the right panel
where $\lambda > m_0$ so that the unproximitized MTI is in a QAH state
at $\lambda'=0$. Proximitized
superconductivity of QAH states is now routinely achieved
experimentally \cite{Chang2013,He2017}.
%{\bf Allan:  Add references to Kang Wang group above.}
%For $\sqrt{\lambda'^2+m_0^2} > |\lambda|$
The QAH state occurs at small $|\mu|$ only when $\sqrt{\lambda'^2+m_0^2} < |\lambda|$;
the gate field $\lambda'$ efficiently converts a QAH insulator with edge states into
a normal insulator with no edge states in the gap, and no opportunity for $Z_2$ states
in ribbons.

%The Chern number $C=2$ BdG state of the
%proximitzed MTI will yield $Z_2=1$ ribbon states at any ribbon width.

% We see that superconductivity
%can
%The ri
%An important observation is that a larger exchange field will
%increase the area of topological regions, and in the QAH regime (when $ \lambda > m_0 $) a broad region of topological phase appears. This will help
%us to design a system to realize MZMs at the ends of quasi-1D regions.
% \begin{figure}
% \centering
% \begin{subfigure}{0.3\linewidth}
% \includegraphics[width=\linewidth]{N_lam1_mu_lam=,5,Del=,2}
% \subcaption{}
% \end{subfigure}
% \begin{subfigure}{0.3\linewidth}
% \includegraphics[width=\linewidth]{N_lam1_mu_lam=1,Del=,2}
% \subcaption{}
% \end{subfigure}
% \begin{subfigure}{0.3\linewidth}
% \includegraphics[width=\linewidth]{N_lam1_mu_lam=2,Del=,2}
% \subcaption{}
% \end{subfigure}
% \caption{Phase Diagrams in $\lambda'$-$\mu$ space, where $\Delta_t=0.2m_0,\Delta_b=0,\lambda=$ (a)$0.5m_0$ (b)$m_0$ (c)$2m_0$.} \label{N_lam1_mu}
% \end{figure}

\begin{figure}
\centering
\begin{subfigure}{0.45\linewidth}
\includegraphics[width=\linewidth]{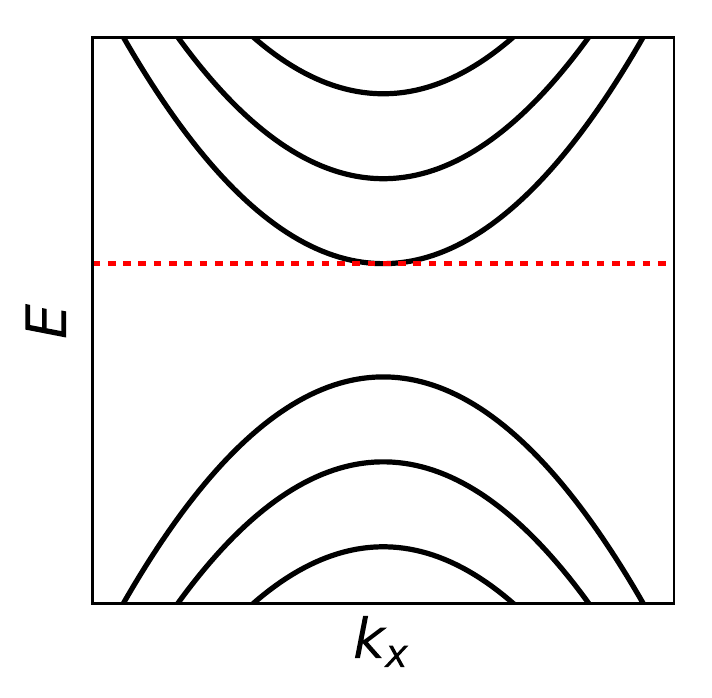}
\subcaption{} \label{subbands}
\end{subfigure}
\begin{subfigure}{0.45\linewidth}
\includegraphics[width=\linewidth]{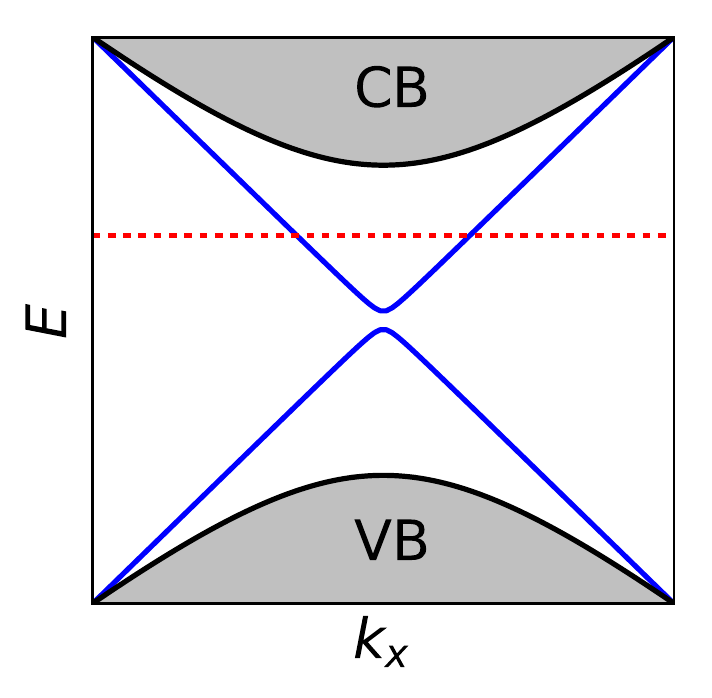}
\subcaption{} \label{qah_band}
\end{subfigure}
\begin{subfigure}{0.45\linewidth}
\includegraphics[width=\linewidth]{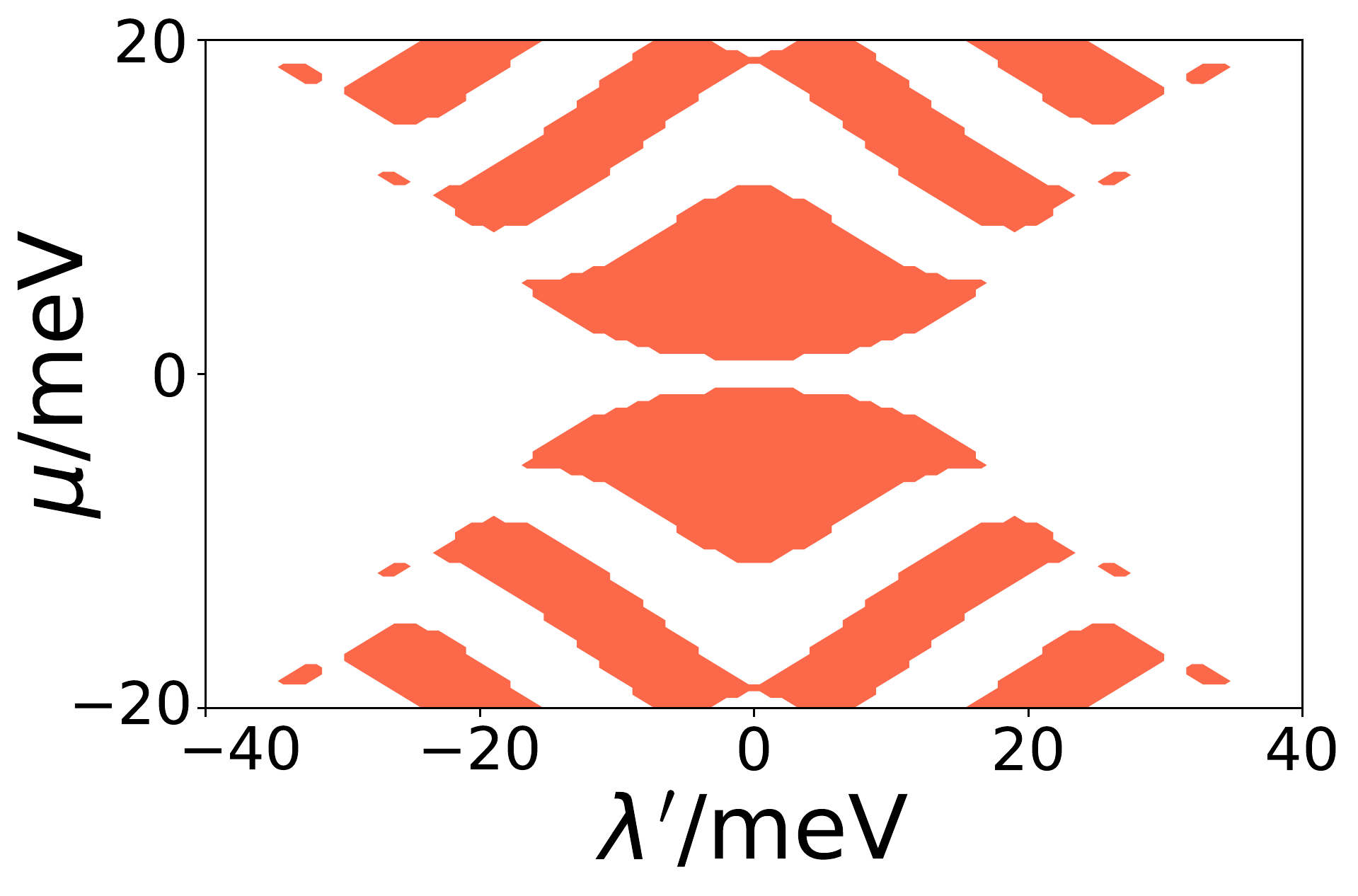}
\subcaption{} \label{phase_w_300}
\end{subfigure}
\begin{subfigure}{0.45\linewidth}
\includegraphics[width=\linewidth]{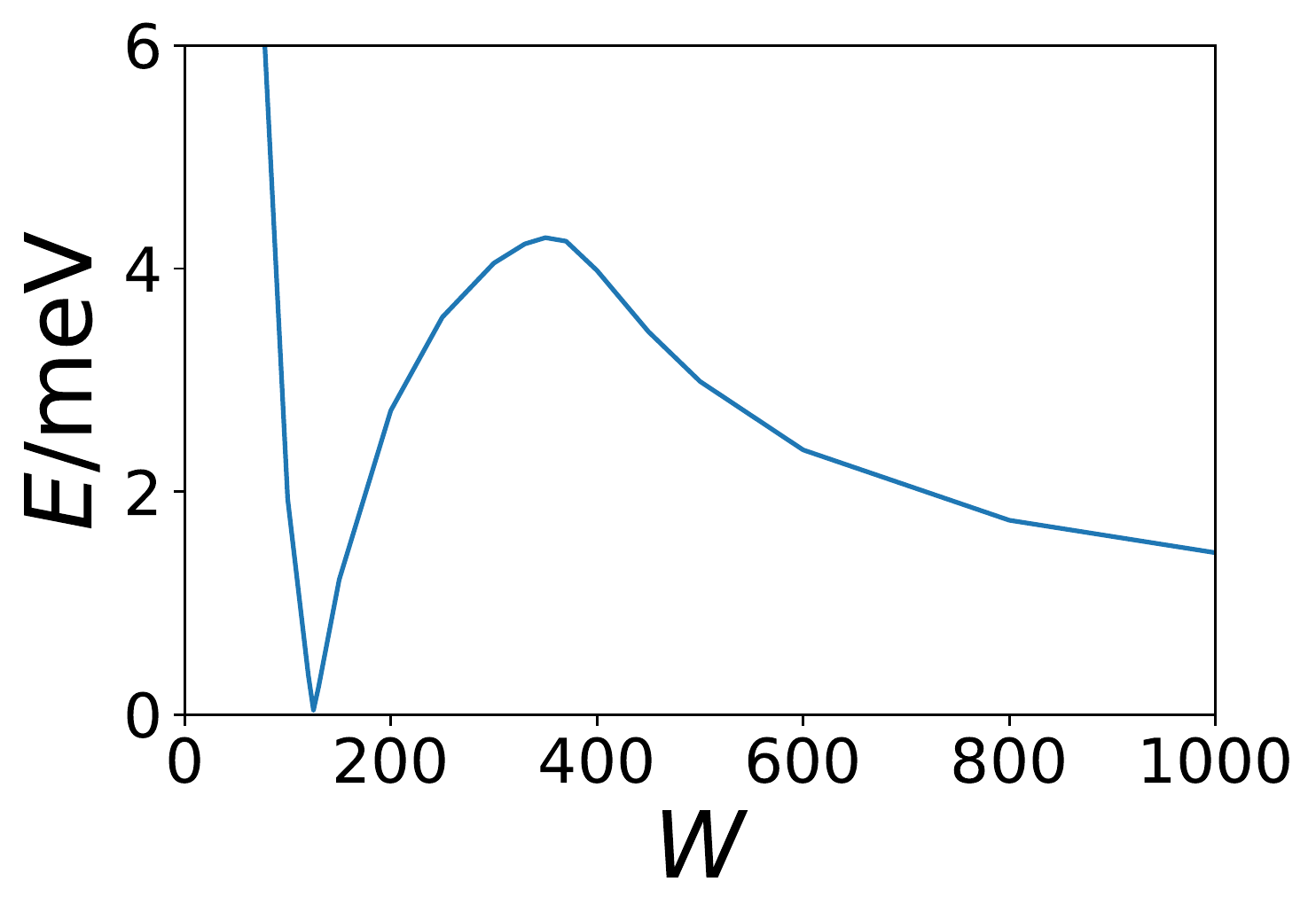}
\subcaption{} \label{gap_width}
\end{subfigure}
\caption{ (a) Bandstructure of a nanoribbon. When the Fermi level (the red dotted line) coincides with the bottom of a subband, a TPT occurs;
(b) Bandstructure in the QAH regime. Grey regions are conduction and valence bands, blue curves are edge states with a small hybridization gap, and the red dotted line shows the Fermi level.
When the chemical potential is inside the bulk gap and outside the hybridization gap, with SC the system should always be nontrivial.
(c) Phase diagram in $\lambda'$-$\mu$ space for a quasi-1D chain with
$W=300, m_0=\SI{6}{\milli\electronvolt}, m_1=\SI{0.2}{\electronvolt}, v=\SI{0.7}{\electronvolt},\lambda=\SI{12}{\milli\electronvolt}, \Delta_b=\SI{1}{\milli\electronvolt}, \Delta_t=0$;
(d) Relation between gap and width, with $\mu=\SI{5}{\milli\electronvolt},\lambda'=0$ and other parameters the same as in (c).} \label{quasi1d_realistic}
\end{figure}

For weak pairing, ribbons have $Z_2=1$ states when an odd number of subbands
cross the Fermi level. In a ribbon formed from
a semiconductor quantum well with strong Zeeman and spin-orbit coupling, or from an
MTI with $\lambda \ll m_0$, a series of closely spaced non-degenerate 1D sub-bands
appear close to the extrema of the
bulk 2D bands, as illustrated schematically in Fig. \ref{subbands}.
When the ribbon width increases, the spacing between subbands becomes smaller
and the $Z_2$ phase diagrams will have more closely spaced boundaries between
$Z_2=0$ and $Z_2=1$ phases. In the $\lambda \gg m_0$  QAH case on the other hand,
illustrated schematically in Fig. \ref{qah_band}, a single
pair of bands crosses the Fermi level at all energies inside the bulk 2D gap,
except for a narrow gapped region due to the avoided crossing between QAH edge states
localized on opposite sides of the ribbon.
The system is therefore almost always nontrivial in the
broad range $\mu\in(-E_b,E_b)$, where $E_b$ is the bulk gap, independent of the ribbon
width.

In Fig. \ref{phase_w_300} we plot the $Z_2$ phase diagram in $(\mu,\lambda')$ plane
of $W=300, m_0=\SI{6}{\milli\electronvolt},
m_1=\SI{0.2}{\electronvolt}, v=\SI{0.7}{\electronvolt}, \lambda=\SI{12}{\milli\electronvolt},
\Delta_b=\SI{1}{\milli\electronvolt}, \Delta_t=0$ ribbon,
which exhibits large adjacent trivial and
nontrivial regions near $\mu=0$.  For these realistic parameters
$|\lambda|$ is larger than $m_0$, but not much larger.  (Larger QAH regions can be achieved by going to
thicker films with smaller $m_0$ but only at the expense of reducing all relevant energy scales.)
QAH edge states overlap strongly even at $W=300$, so strongly in fact that the
normal state gap produced by avoided crossing of edge states is comparable to $m_0$,
creating a substantial $Z_2=0$ region near $\mu=0$.  This region is however still
bordered at $\lambda'=0$ by a large $Z_2=1$ region that can be identified with pairing of
QAH edge states. The gate field $\lambda'$ sweeps this state into a
large adjacent $Z_2=0$ region, which can be identified with a proximity
coupled ordinary insulator.

Fig. \ref{gap_width} plots the ribbon width dependence of the
gap at $k_x=0$,  $\lambda'=0$, and $\mu=\SI{5}{\milli\electronvolt}$ out to
$W \sim 1000$ in lattice constant units.
Here we notice that there is only one gap closing as a function of $W$ which
signals a phase transition between a small $W$ $Z_2=0$ state
associated with a large avoided crossing gap between QAH edge states,
and a large $W$ $Z_2=1$ state associated with pairing of
QAH edge states. The gaps remain large out to
$\sim 1000$ lattice constants, corresponding
to a physical length $\sim \SI{400}{\nano\meter}$,
partly because the edge state localization length is enhanced by the relatively high
velocity of quantum Hall edge states compared to the velocity of states
present near the bottom of a bulk 2D band.
As the system becomes wider, the QAH edge states have
less overlap and the gap is eventually reduced to very small values.
%When $W=145$ the bottom of QAH gap goes down to the Fermi level
%$\mu=\SI{5}{\milli\electronvolt}$, which corresponds to a gap closing in the SC spectrum, so the system enters the nontrivial regime. When $W$ further increases, since $\mu$ is inside the bulk gap, the system is always nontrivial, and at large $W$ the gap decreases exponentially. Such a robust quasi-1D topological phase compared to the other quasi-1D systems, where topological phase is highly sensitive to change in width, makes this system particularly special to realise MZM.

%\begin{figure}
%\begin{subfigure}{0.45\linewidth}
%\includegraphics[width=\linewidth]{quasi1d_real_w=300}
%\subcaption{}
%\end{subfigure}
%\begin{subfigure}{0.45\linewidth}
%\includegraphics[width=\linewidth]{N_lam1_mu_real}
%\subcaption{}
%\end{subfigure}
%\caption{(a) Phase diagram in $\lambda'$-$\mu$ space for a quasi-1D chain with
%$W=300, m_0=\SI{6}{\milli\electronvolt}, m_1=\SI{0.2}{\electronvolt}, v=\SI{0.7}{\electronvolt},\lambda=\SI{10}{\milli\electronvolt}, \Delta_t=\SI{1}{\milli\electronvolt}, \Delta_b=0$;
%(b) Phase diagram for a 2D system with the same parameters.} \label{quasi1d_realistic}
%\end{figure}

%\item Next Section:  Topological surrounded by Trivial Gives Majoranas
%\textit{Majorana states in topological region surrounded by trivial ones}
%\section{Experiment Scheme}

Because of the bulk-edge correspondence of topological states we expect Majorana zero modes to
appear not only in ribbons, but also in wide films in which quasi-one-dimensional regions are formed that
have local model parameters in the topologically non-trivial range.
In order to define a quasi-1D region, we can form a gate array above a proximitized
MTI thin film whose chemical potential is in the QAH gap, as illustrated schematically in Fig.~\ref{setup}.
Just as in the ribbon case, the gate electric field $\lambda'$ will form local regions
that have $Z_2=0$ and Chern number $N=0$, isolating the $Z_2=1$ and $N=1$ or 2 regions along the less strongly disturbed portions of the
MTI thin film. As discussed in the Supplemental Material \cite{supplement}, we have explicitly verified
that MZMs appear near the ends of quasi-1D topological superconductors that are
written onto the MTI surface in this way.

\noindent
\textit{Conclusion:}
%which is also very convenient for building controlled devices under the currently proposed braiding schemes.
%This strategy for generating MZM resources for
%quantum information processing has obvious advantages compared to
%arrays of separately grown one-dimensional quantum wires, which may be difficult to
%align.  %Out proposed platform with the possibility of manipulating the Majorana states with electric means and
%growing gates with top-down fabrication will pave a way to make larger MZMs networks without external magnetic fields.
Topological quantum computation requires flexible Majorana braiding
that relies on branched structures like T-junctions \cite{Alicea2011}.
Although conceptually simple, T-junctions based on semiconductor quantum wires
are difficult to build because of challenges in depositing aligned semiconductor quantum wires.
We have demonstrated the possibility of
using gate arrays to write MZMs onto the surface of a 2D MTI
placed on a superconducting substrate.
It has some similarities with systems \cite{Pientka2017,Hell2017} in which a gate array
writes quantum wires onto a quantum well by periodically depleting
all carriers, or varying the number of locally occupied subbands between even and
odd values, but has advantages in this case as well because
$\it{i:}$ it is not necessary to apply a magnetic field to break
degeneracies at time-reversal invariant points in one-dimensional momentum space
and because $\it{ii:}$ there can be a large energy separation between
the quasi-1D bands formed by quantum Hall edge states and higher energy subbands,
providing a large target for efforts to tune to $Z_2 = 1$ superconductors.
Additionally, our proposal also provides an ideal platform for building the Majorna box qubits
recently proposed in Refs.~\onlinecite{Plugge2017,Karzig2017}
because $\it{i:}$ it is easy to define arbitrary number of parallel gate-controlled quasi-1D TSC wires as shown in Fig.~\ref{MajoranaSetup}, and $\it{ii:}$ because large topological stability ranges
allows geometrical capacitance\cite{Karzig2017} to be changed without changing topological states.
%MZMs localized at quantum wire ends are likely to prove more amenable to
%electrical control than MZMs localized at the vortices of

This work was supported by the Army Research Office (ARO)
under contract W911NF-15-1-0561:P00001, and by the Welch Foundation
under grant TBF1473.
%\bibliography{bib_majo}

%\begin{document}
\newpage
\begin{widetext}
\section{Supplementary Material for\\ A Quantum Anomalous Hall Majorana Platform}
\end{widetext}

\section{Material Parameters}
\begin{figure}
\centering	
\begin{subfigure}[b]{0.45\linewidth}
\includegraphics[width=\linewidth]{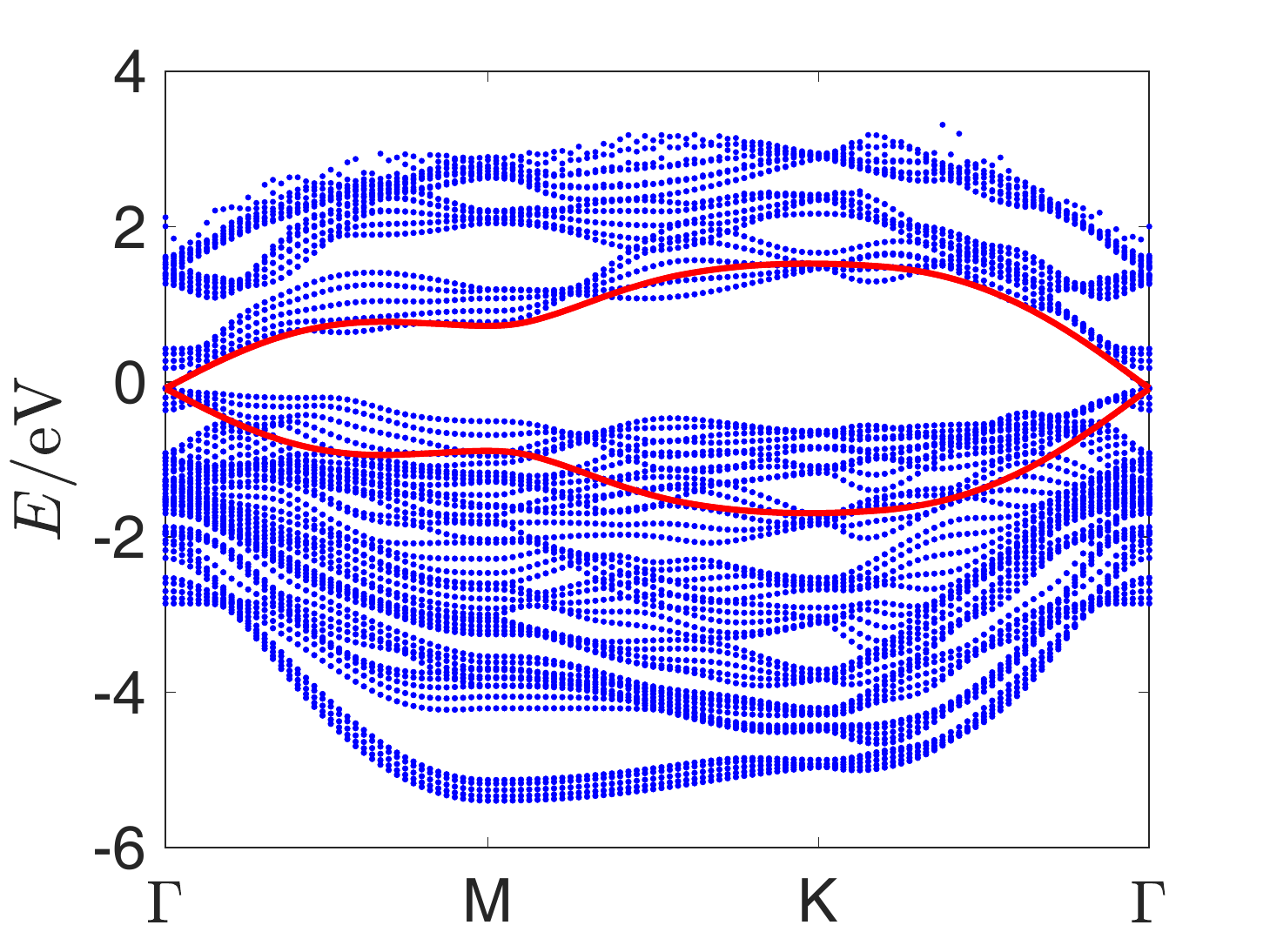}
\caption{} \label{bi2se3_band_u}
\end{subfigure}
\begin{subfigure}[b]{0.45\linewidth}
\includegraphics[width=\linewidth]{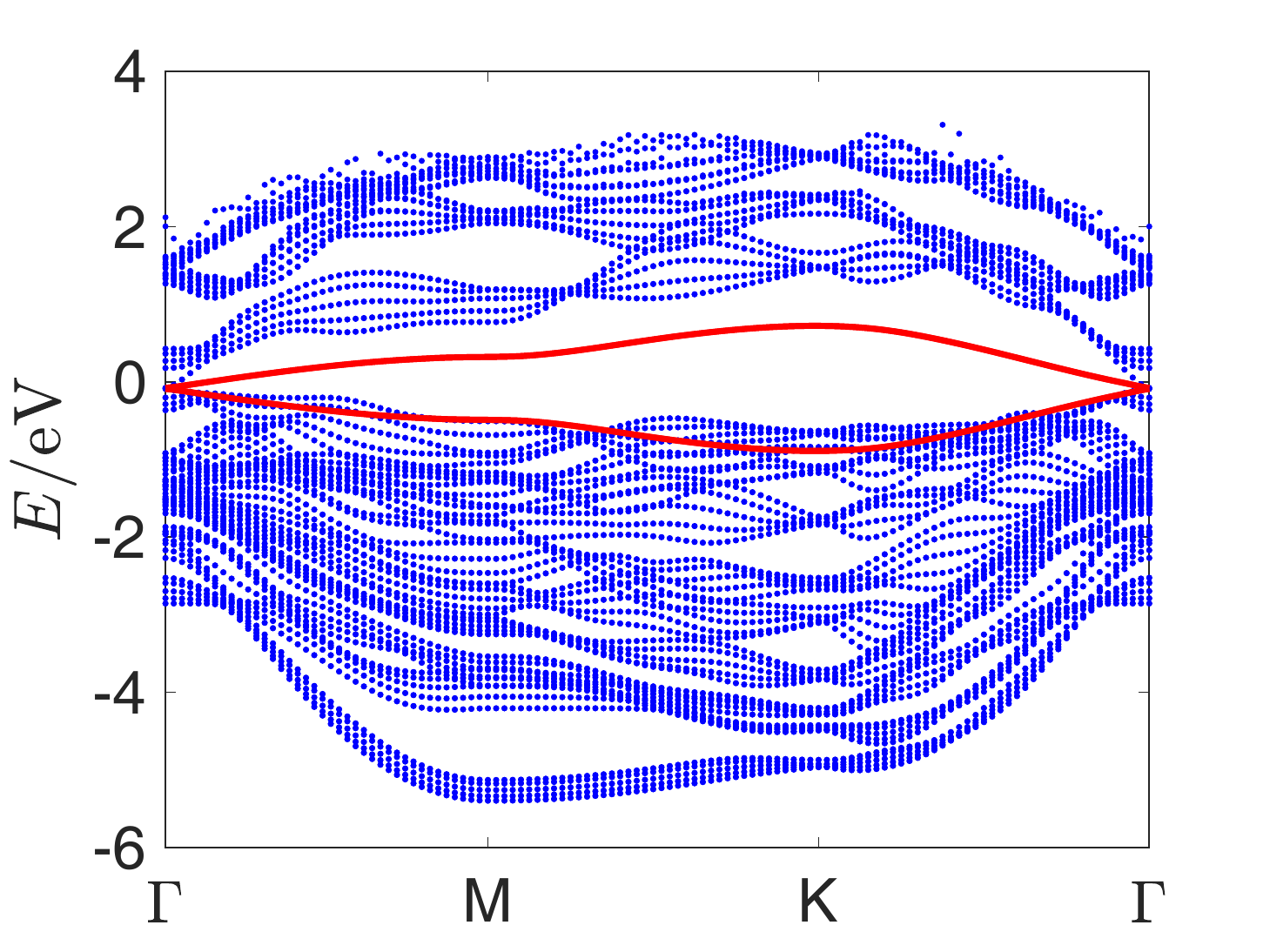}
\caption{} \label{bi2se3_band_d}
\end{subfigure}
\caption{Bandstructure of a 5-layer Bi$_2$Se$_3$ thin film. Blue dots are from DFT calculation,
while red curves are from our effective lattice model when we choose $m_0=\SI{6}{\milli\electronvolt}$ and (a) $m_1=\SI{0.2}{\electronvolt}$, $v=\SI{0.7}{\electronvolt}$; (b) $m_1=\SI{0.1}{\electronvolt}$, $v=\SI{0.2}{\electronvolt}$.} \label{bi2se3_band}
\end{figure}

Density functional theory performed with Vienna Ab initio simulation package (VASP) is used to calculate the electronic structure of 5-layer Bi$_2$Se$_3$, and the bandstructure is shown
in Fig. \ref{bi2se3_band}. The bandstructure from the lattice model is also plotted, with a rough estimate of the hybridization
of two surfaces $m_0=\SI{6}{\milli\electronvolt},m_1=\SI{0.2}{\electronvolt}$ and Fermi velocity $v=\SI{0.7}{\electronvolt}$ in Fig. \ref{bi2se3_band_u},
and $m_0=\SI{6}{\milli\electronvolt}, m_1=\SI{0.1}{\electronvolt}, v=\SI{0.2}{\electronvolt}$ in Fig. \ref{bi2se3_band_d}, which are two limits
with the parameters fitting well with the surface band above and below the Fermi level).
The lattice model has good agreement with DFT results and experiments.

\section{Lattice Model}
With substitutions $ k_{x/y} \to \sin k_{x/y} $ and
$ k^2 \to 2(2-\cos k_x - \cos k_y) $,
we can rewrite the momentum-space Hamiltonian in the following form
\begin{equation}
H_0(\bm k) = h_0 + (h_xe^{ik_x}+h_ye^{ik_y}+h.c.)
\end{equation}
Then we obtain a nearest-neighbor tight-binding model, where the onsite term is
\begin{equation}
h_0 = (m_0+4m_1)\tau_x + \lambda\sigma_z + \lambda'\tau_z
\end{equation}
and the hopping terms in $x$ and $y$ directions are
\begin{align}
h_x &= -\frac i2 v\sigma_x - m_1\tau_x \\
h_y &= \frac i2 v\sigma_y - m_1\tau_x
\end{align}
With $s$-wave SC, the on-site and hopping matrices become
\begin{align}
h_0^{\textrm{sc}} &= \left(\begin{array}{cc}
h_0-\mu & \Delta_\textrm{sc} \\
\Delta_\textrm{sc}^\dagger & -h_0^*+\mu
\end{array}\right)\\
h_x^{\textrm{sc}} &= \left(\begin{array}{cc}
h_x & 0 \\ 0 & -h_x^*
\end{array}\right)\\
h_y^{\textrm{sc}} &= \left(\begin{array}{cc}
h_y & 0 \\ 0 & -h_y^*
\end{array}\right)
\end{align}
\section{Block Diagonalization of the BdG Matrix\label{App:BlockDiag}}
\begin{figure}
\centering	
\begin{subfigure}[b]{0.52\linewidth}
\includegraphics[width=\linewidth]{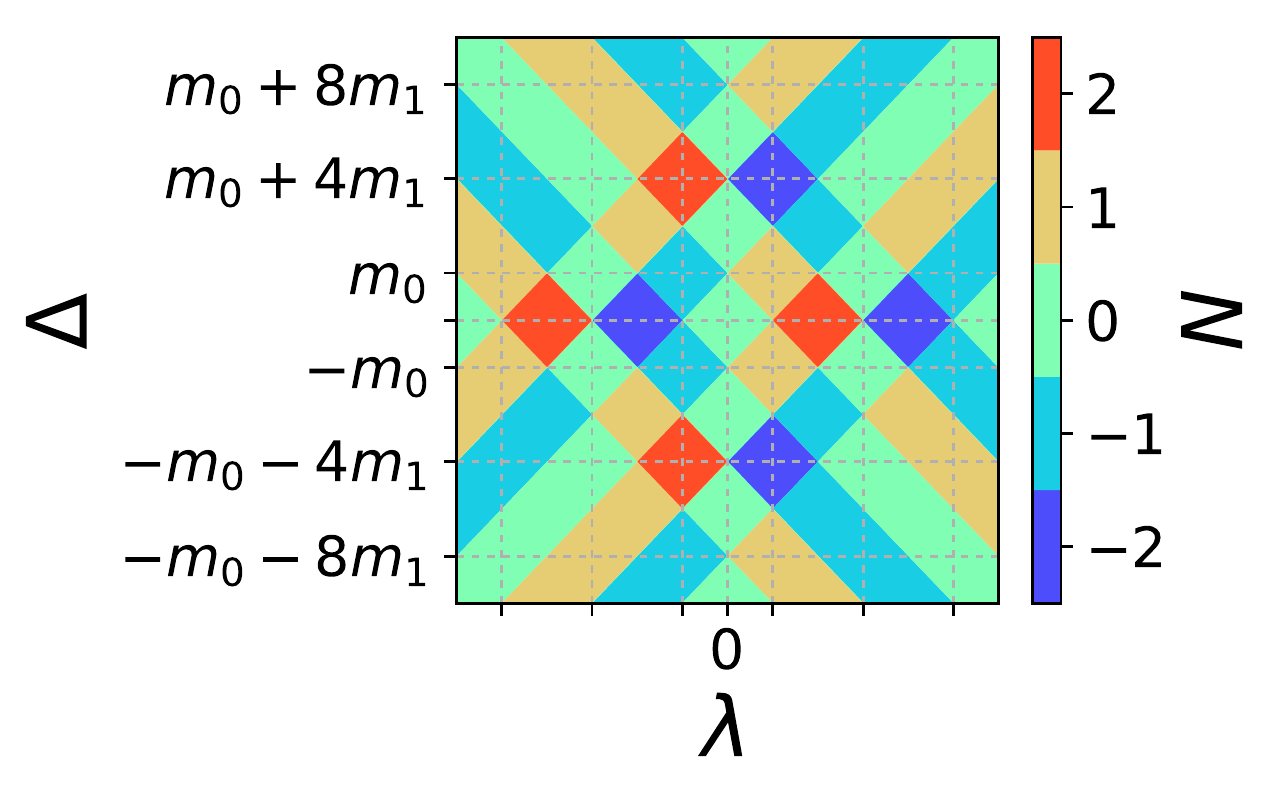}
\caption{} \label{N_lam_Del_lattice}
\end{subfigure}
\begin{subfigure}[b]{0.45\linewidth}
\includegraphics[width=\linewidth]{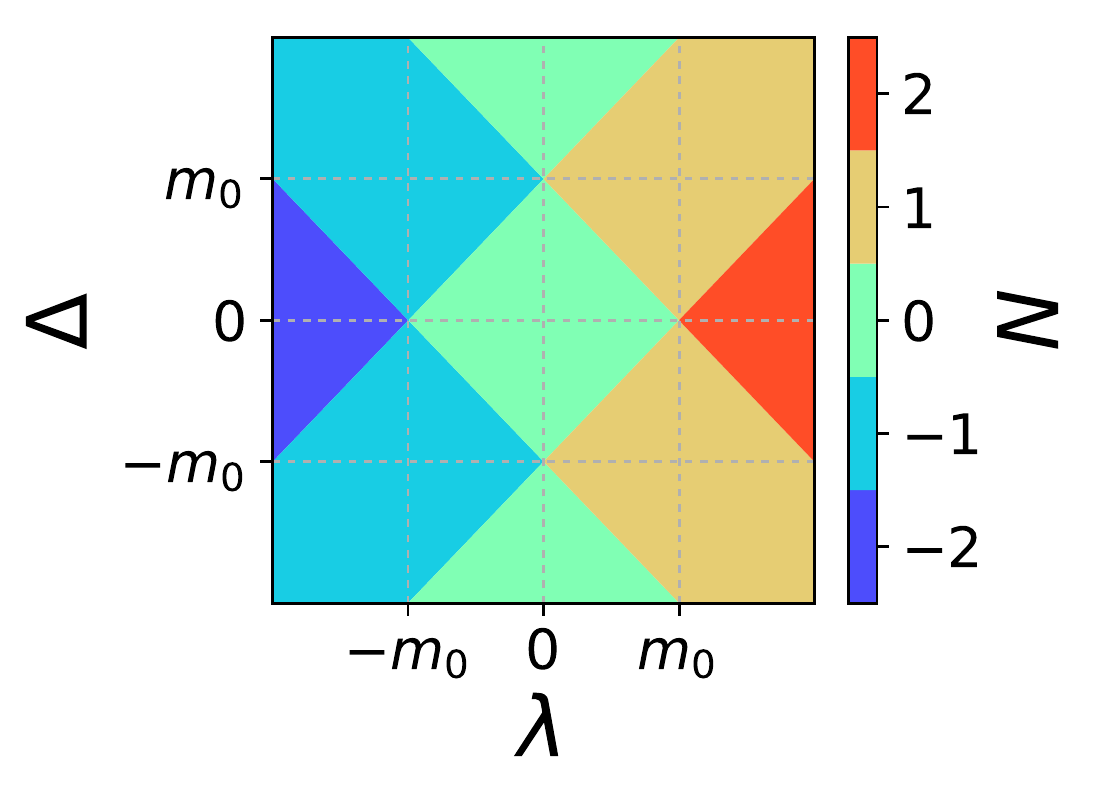}
\caption{} \label{N_lam_Del_continuum}
\end{subfigure}
\caption{Lattice (a) and continuum (b) model Chern number $N$ phase diagrams for
2D geometries with $\lambda'=\mu=0$ and
$\Delta_t=-\Delta_b=\Delta$.  For visualization purposes
(a) is plotted at an unphysically small value of $m_1/m_0$.}
\end{figure}

In the new basis $\phi_{\bm k}=U\psi_{\bm k}$, where
\begin{equation}
U = \frac 12 \left(\begin{array}{cccccccc}
1 & 0 & 1 & 0 & 0 & 1 & 0 & -1 \\
0 & 1 & 0 & -1 & 1 & 0 & 1 & 0 \\
0 & -1 & 0 & 1 & 1 & 0 & 1 & 0 \\
1 & 0 & 1 & 0 & 0 & -1 & 0 & 1 \\
0 & 1 & 0 & 1 & 1 & 0 & -1 & 0 \\
1 & 0 & -1 & 0 & 0 & 1 & 0 & 1 \\
-1 & 0 & 1 & 0 & 0 & 1 & 0 & 1 \\
0 & 1 & 0 & 1 & -1 & 0 & 1 & 0
\end{array}\right)
\end{equation}
the BdG matrix can be block-diagonalized
\begin{equation}
H_\textrm{BdG}(\bm k) = \textrm{diag}\{\bm{d}_1\cdot\bm{\sigma},\bm{d}_2\cdot\bm{\sigma},\bm{d}_3\cdot\bm{\sigma},\bm{d}_4\cdot\bm{\sigma}\}
\label{eq:H_BdG_Diag}
\end{equation}
where
\begin{align}
\bm{d}_1 = &(v\sin k_y, -v\sin k_x, m_0+\lambda+\Delta\notag\\&+2m_1(2-\cos k_x-\cos k_y)) \notag\\
\bm{d}_2 = &(-v\sin k_y, -v\sin k_x, -m_0-\lambda+\Delta\notag\\&-2m_1(2-\cos k_x-\cos k_y)) \notag\\
\bm{d}_3 = &(v\sin k_y, v\sin k_x, m_0-\lambda-\Delta\notag\\&+2m_1(2-\cos k_x-\cos k_y))\notag\\
\bm{d}_4 = &(-v\sin k_y, v\sin k_x, -m_0+\lambda-\Delta\notag\\&-2m_1(2-\cos k_x-\cos k_y))
\end{align}
For each block the Chern number can be calculated with the formula
\begin{equation}
N_i = \frac{1}{4\pi}\int_{\textrm{BZ}} d^2k\, \hat{\bm{d}_i}\cdot(\partial_{k_x}\hat{\bm{d}_i}\times\partial_{k_y}\hat{\bm{d}_i})
\end{equation}
Then the Chern number of the whole system is $N = \sum_{i=1}^{4} N_i$. The phase diagram is shown in Fig. \ref{N_lam_Del_lattice}.
\par A general $p\pm ip$ TSC in 2D is described by Hamiltonian $H_p = \bm{h}\cdot\bm{\sigma}$, with $\bm{h} = (\Delta_p\sin k_x, \pm \Delta_p\sin k_y, 2t(2-\cos k_x-\cos k_y)-\mu)$. In its non-trivial phase $|\mu| < 2t$ the chiral Majorana mode appear, which is localized in the $\xi_P\sim \frac{\Delta}{|\mu|}$ width near the edge. Comparing $\bm{h}$ with $\bm{d_i}$s, we get edge mode localization lengths  for effective $p\pm ip$ superconductors of \eqref{eq:H_BdG_Diag}.
\begin{align}
&\xi_1 = \frac{v}{|m_0+\lambda+\Delta|}\quad
\xi_2 = \frac{v}{|-m_0-\lambda+\Delta|}\notag\\
&\xi_3 = \frac{v}{|m_0-\lambda-\Delta|}\quad
\xi_4 = \frac{v}{|-m_0+\lambda-\Delta|}
\label{Eq:LocalizationLength}
\end{align}

\par
\begin{figure}
\centering
\begin{subfigure}{0.45\linewidth}
\includegraphics[width=\linewidth]{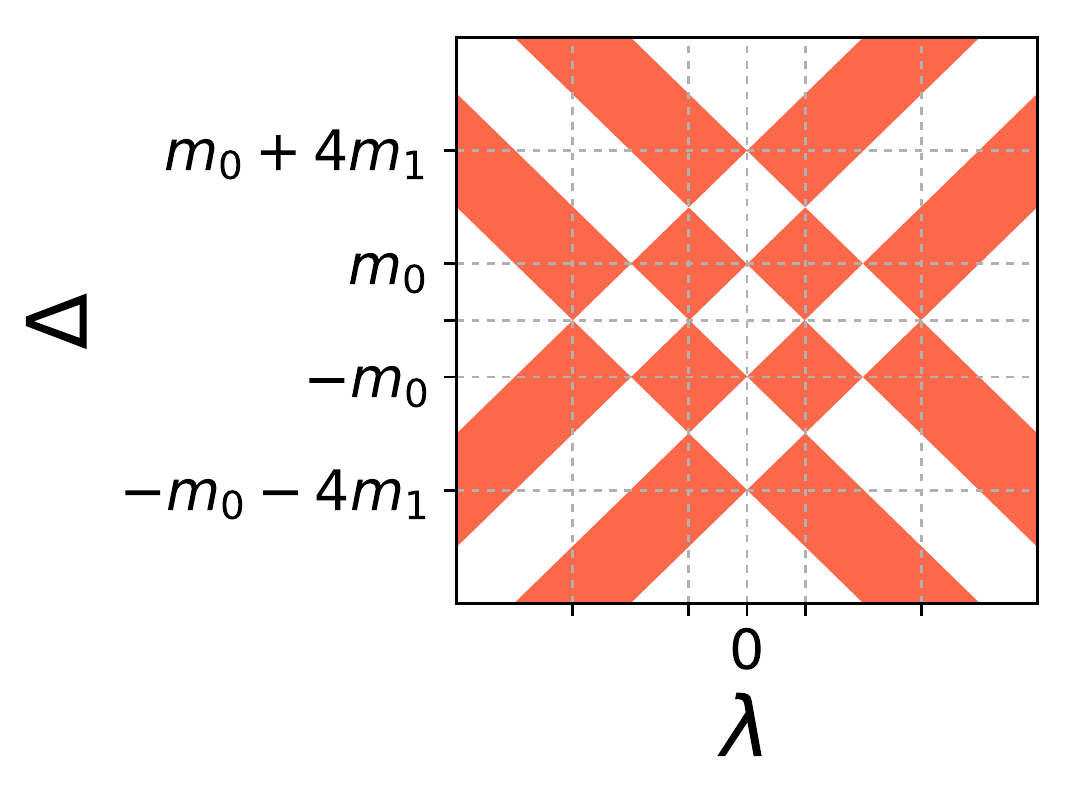}
\subcaption{} \label{quasi1d_w=1}
\end{subfigure}
\begin{subfigure}{0.45\linewidth}
\includegraphics[width=\linewidth]{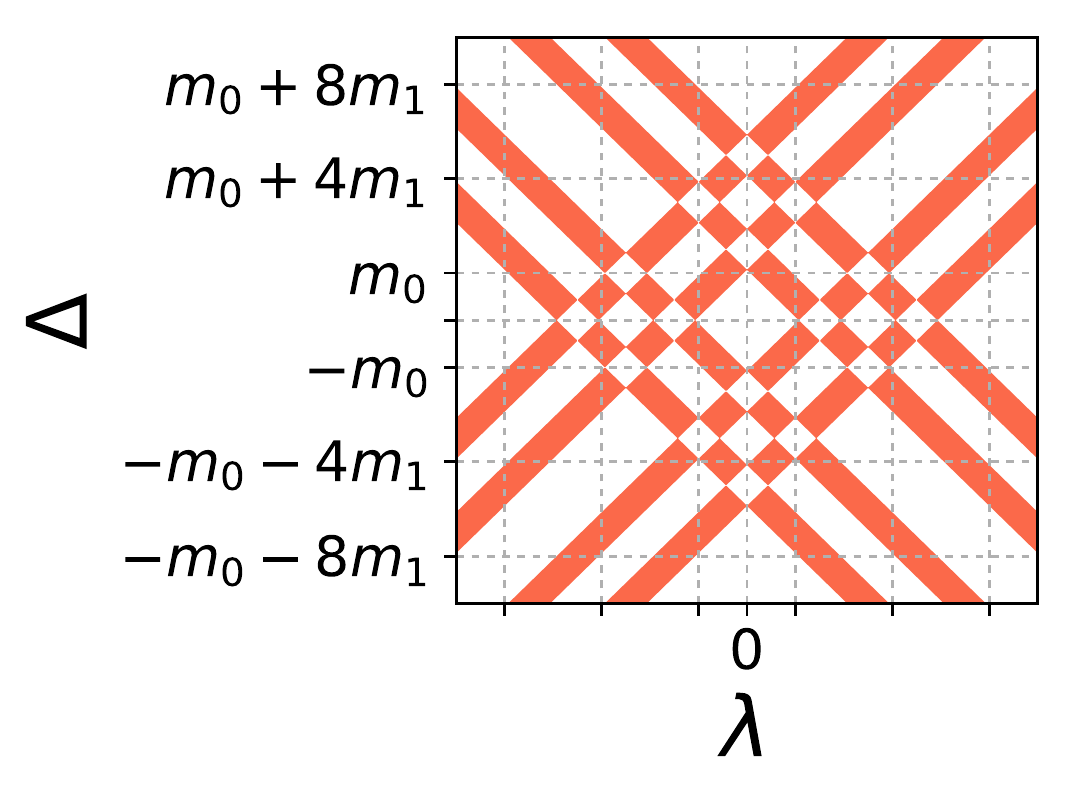}
\subcaption{}
\end{subfigure}
\begin{subfigure}{0.45\linewidth}
\includegraphics[width=\linewidth]{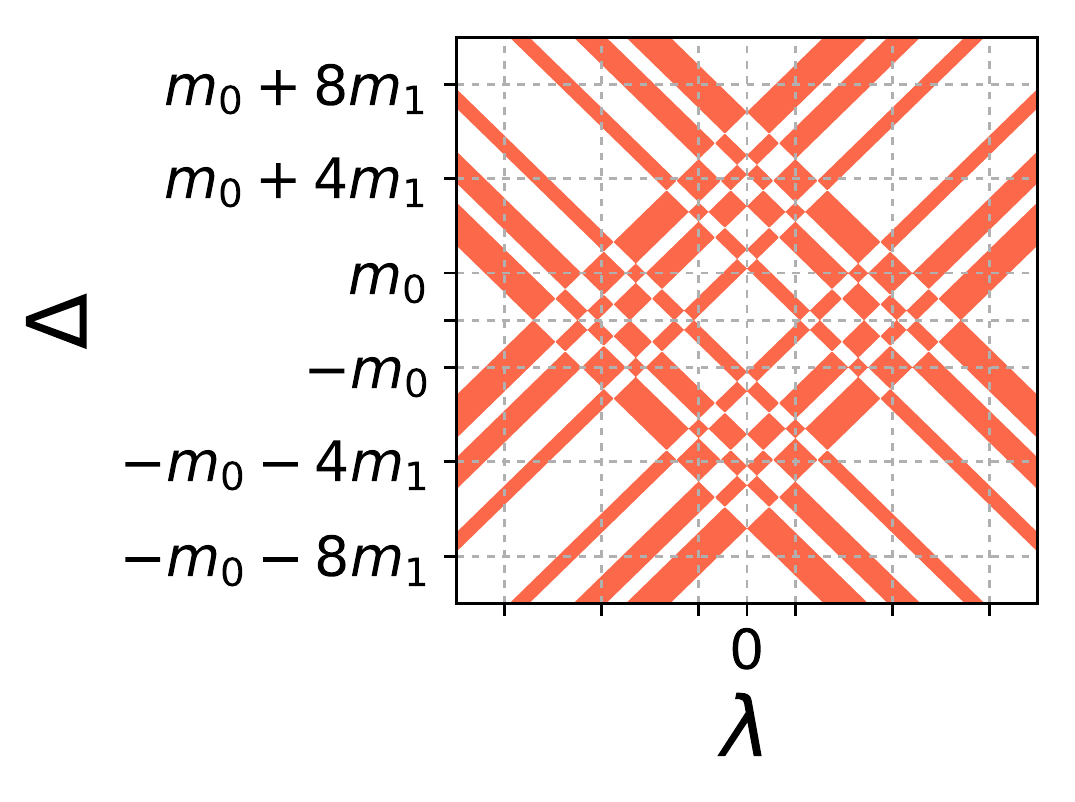}
\subcaption{}
\end{subfigure}
\begin{subfigure}{0.45\linewidth}
\includegraphics[width=\linewidth]{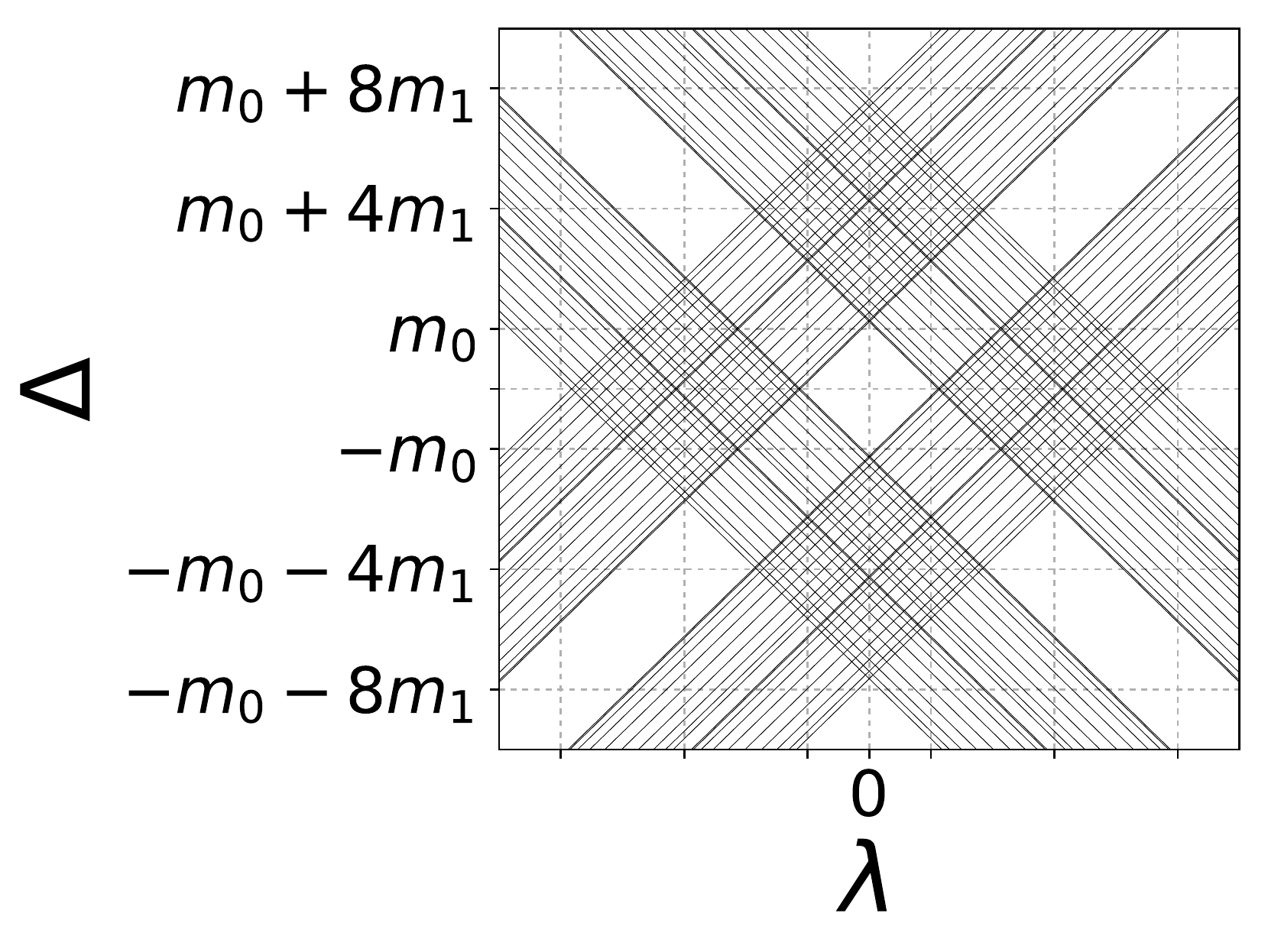}
\subcaption{} \label{quasi1d_w=10}
\end{subfigure}
\caption{Phase diagrams of (quasi-)1D systems with $\lambda'=\mu=0,\Delta_t=-\Delta_b=\Delta$, and
chain width $W=$ (a)1 (b)2 (c)3 (d)10.
In (a)(b)(c), red and white regions are respectively
topologically nontrivial and trivial.  The phase diagram will vary continuously as $\Delta_{t,b}$ and other
model parameters are varied.
Only the phase boundaries are plotted in (d).  For visualization purposes
these figures are plotted for an unphysically small value of $m_1/m_0$.
} \label{quasi1d_phase}
\end{figure}

 For a 1D wire, $y$-direction hopping is ignored, so the effective Hamiltonian for 3DTI surface states becomes
$h_{\textrm{eff}}(k_x) = -\sin k_x\sigma_y$, and the hybridization term becomes $m_k=m_0+2m_1(1-\cos k_x)$.
If we still assume $\lambda'=\mu=0,\Delta_t=-\Delta_b=\Delta$, then when we perform the same basis transformation as above, the BdG matrix will be block-diagonalized into four $2\times2$ blocks.
If we consider the first block, which is (we write $k_x$ as $k$ for simplicity)
\begin{equation}
h_1(k) = \left(\begin{array}{cc}
m_k+\lambda+\Delta & iv\sin k \\
-iv\sin k & -m_k-\lambda-\Delta
\end{array}\right)
\end{equation}
with the basis $(\alpha_k,\alpha_{-k}^\dagger)^T$ where $ \alpha_k = c_{k\uparrow}^t + c_{k\uparrow}^b + c_{-k\downarrow}^{t\dagger} - c_{-k\downarrow}^{b\dagger} $.
%\begin{equation}
%\alpha_k = c_{k\uparrow}^t + c_{k\uparrow}^b + c_{-k\downarrow}^{t\dagger} - c_{-k\downarrow}^{b\dagger}
%\end{equation}
For a wire with $N$ lattice sites, the Hamiltonian can be written in real space:
\begin{equation}
\begin{split}
h_1 = &(m_0+2m_1+\lambda+\Delta)\sum_{j=1}^{N}\alpha_j^\dagger\alpha_j \\&+ \sum_{j=1}^{N-1}(m_1\alpha_j^\dagger\alpha_{j+1}-\frac v2 \alpha_j\alpha_{j+1} + h.c.)
\end{split}
\end{equation}
This is in the same form as a Kitaev chain \cite{Kitaev2001}. When $|m_0+2m_1+\lambda+\Delta|<2m_1$, Majorana fermions will appear at the ends. When we consider all the four blocks, the resulting phase diagram is shown in Fig. \ref{quasi1d_w=1}.

For a quasi-1D system with larger width $W$, the Hamiltonian is an $8W\times 8W$ matrix. It can still be diagonalized into four blocks, each of size $2W\times2W$. The phase boundaries cannot be solved analytically, but we can prove that they are still straight lines. For $W=1,2,3$ and 10, the phase diagrams are shown in Fig. \ref{quasi1d_phase}. When $W$ increases, more phase boundaries appear, resulting in an alternating pattern in the phase diagrams. When $W$ is large enough, the outline of the pattern will look very similar to that in Fig. \ref{N_lam_Del_lattice}.

\section{$Z_2$ Invariant}
The $Z_2$ invariant of a 1D system can be obtained by Pfaffian calculation. For a BdG matrix in the form
\begin{equation}
H_{\textrm{BdG}}(k) = \left(\begin{array}{cc}
h(k) & \Delta \\
\Delta^{\dagger} & -h^*(-k)
\end{array}\right)
\end{equation}
where $h(k)$ and $\Delta$ are matrices, a skew-symmetrized matrix can be obtained in Majorana basis:
\begin{equation}\label{majorana_basis}
  \left(\begin{array}{cc}
\gamma_1  \\
\gamma_2
\end{array}\right)
 = \frac{1}{\sqrt{2}}\left(\begin{array}{cc}
1 & 1 \\
-i & i
\end{array}\right)
 \left(\begin{array}{cc}
c^{\dagger}  \\
c
\end{array}\right)
\end{equation}
with $\gamma_i$ is the creation operator of Majorana Fermion and $c^{\dagger}/c$ is the creation/annihilation operator of electrons.
Then we can get the skew-symmetrized Hamiltonian as:
\begin{equation}
\tilde{H}(k) = U H_{\textrm{BdG}}(k) U^{\dagger}
\end{equation}
where
\begin{equation}
U = \frac{1}{\sqrt{2}}\left(\begin{array}{cc}
1 & 1 \\
-i & i
\end{array}\right)
\end{equation}
where the numbers denote matrices proportional to the identity matrix with the same dimension as $h(k)$. Then
\begin{equation}
\begin{split}
&\tilde{H}(k) = \\
&{\footnotesize \left(\begin{array}{cc}
h(k)-h^*(-k)+\Delta+\Delta^\dagger & i(h(k)+h^*(-k)-\Delta+\Delta^\dagger) \\
-i(h(k)+h^*(-k)+\Delta-\Delta^\dagger) & h(k)-h^*(-k)-\Delta-\Delta^\dagger
\end{array}\right)}
\end{split}
\end{equation}
Since $h(k)=h(-k)=h^\dagger(-k), \Delta^\dagger=-\Delta^*$, $\tilde{H}(k)$ is a skew-symmetric matrix. The $Z_2$ invariant can then be evaluated as $ \nu = \textrm{sgn}[\textrm{Pf}\tilde{H}(0)] \textrm{sgn}[\textrm{Pf}\tilde{H}(\pi)]$, where Pf denotes the Pfaffian number and $\textrm{sgn}[x]$ is the sign of $x$. Numerical calculation of the Pfaffian number is performed using the Pfapack code \cite{wimmer2012algorithm}.

\section{Fitting the Lower Band}
In the main text, all calculations are based on parameters from Fig. \ref{bi2se3_band_u}, which fit the upper band pretty well. Here we present results from calculations with much smaller $v$ and $m_1$ ($v=\SI{0.2}{meV},m_1=\SI{0.1}{meV}$), which give better fitting of the surface band below the Fermi level as shown in Fig. \ref{bi2se3_band_d}.
\begin{figure*}
	\centering	
	\begin{subfigure}[b]{0.25\linewidth}
		\includegraphics[width=\linewidth]{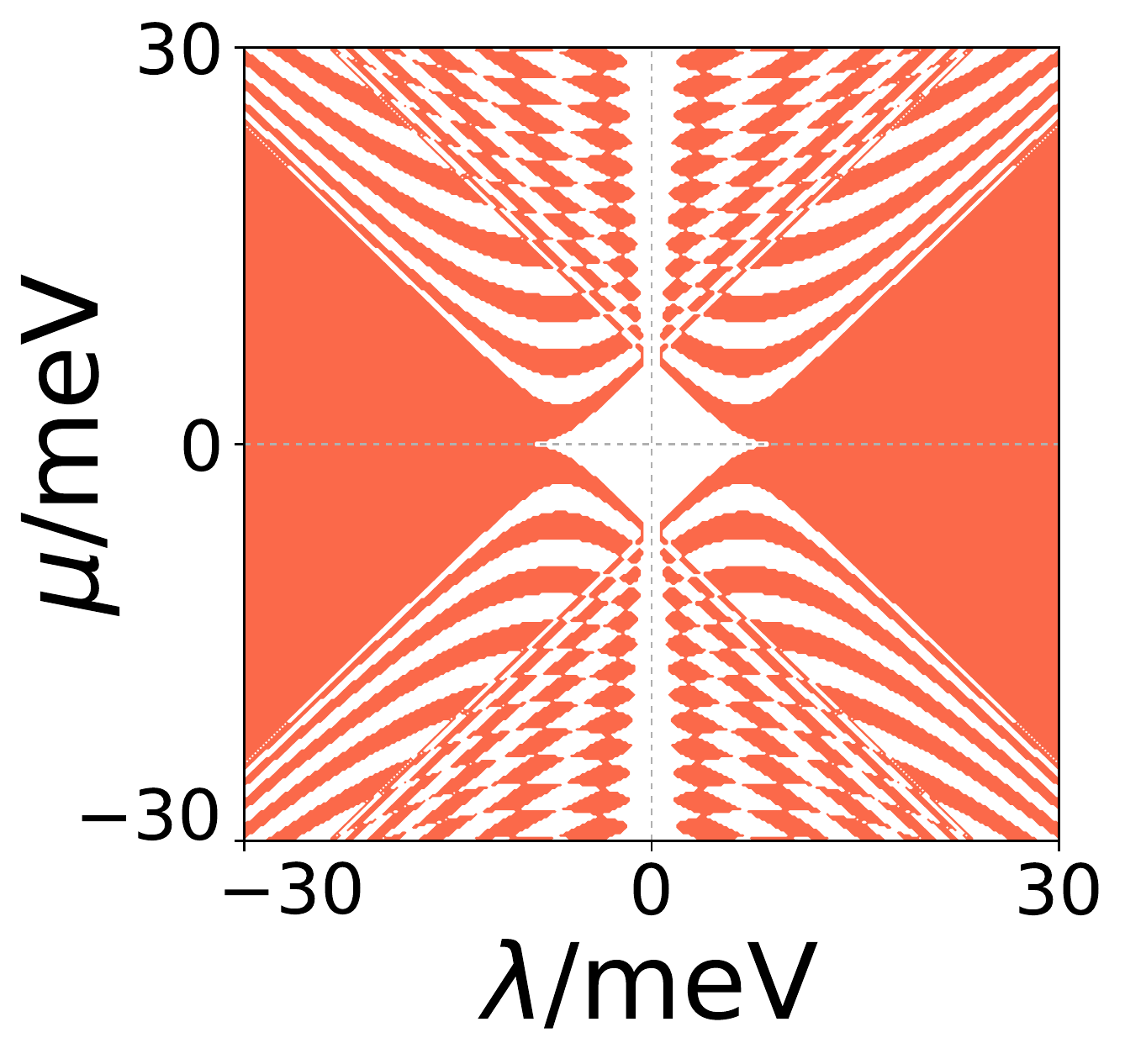}
		\caption{} \label{phase_lam_mu_v}
	\end{subfigure}
	\begin{subfigure}[b]{0.35\linewidth}
		\includegraphics[width=\linewidth]{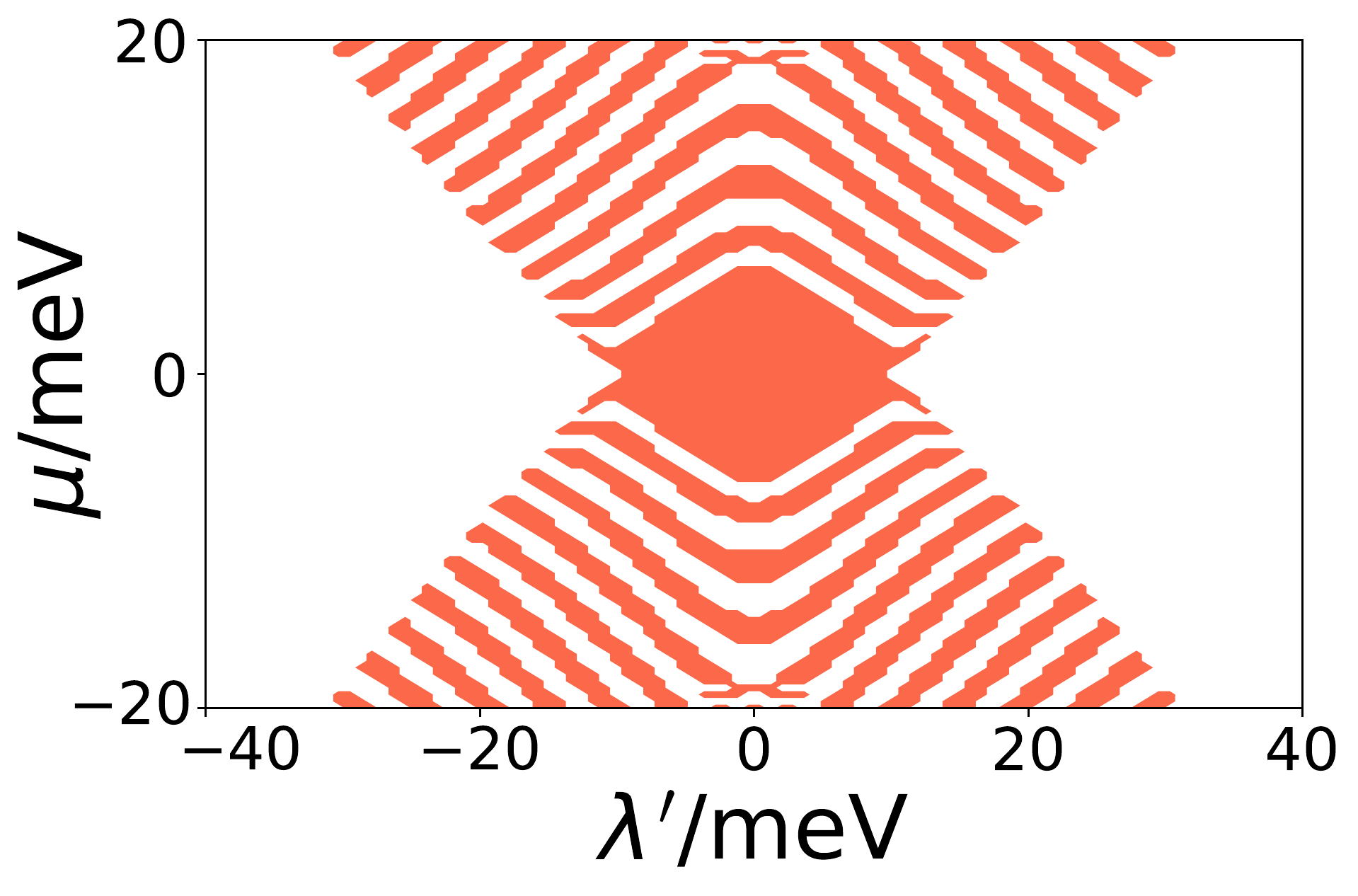}
		\caption{} \label{phase_w_300_v}
	\end{subfigure}
	\begin{subfigure}[b]{0.33\linewidth}
		\includegraphics[width=\linewidth]{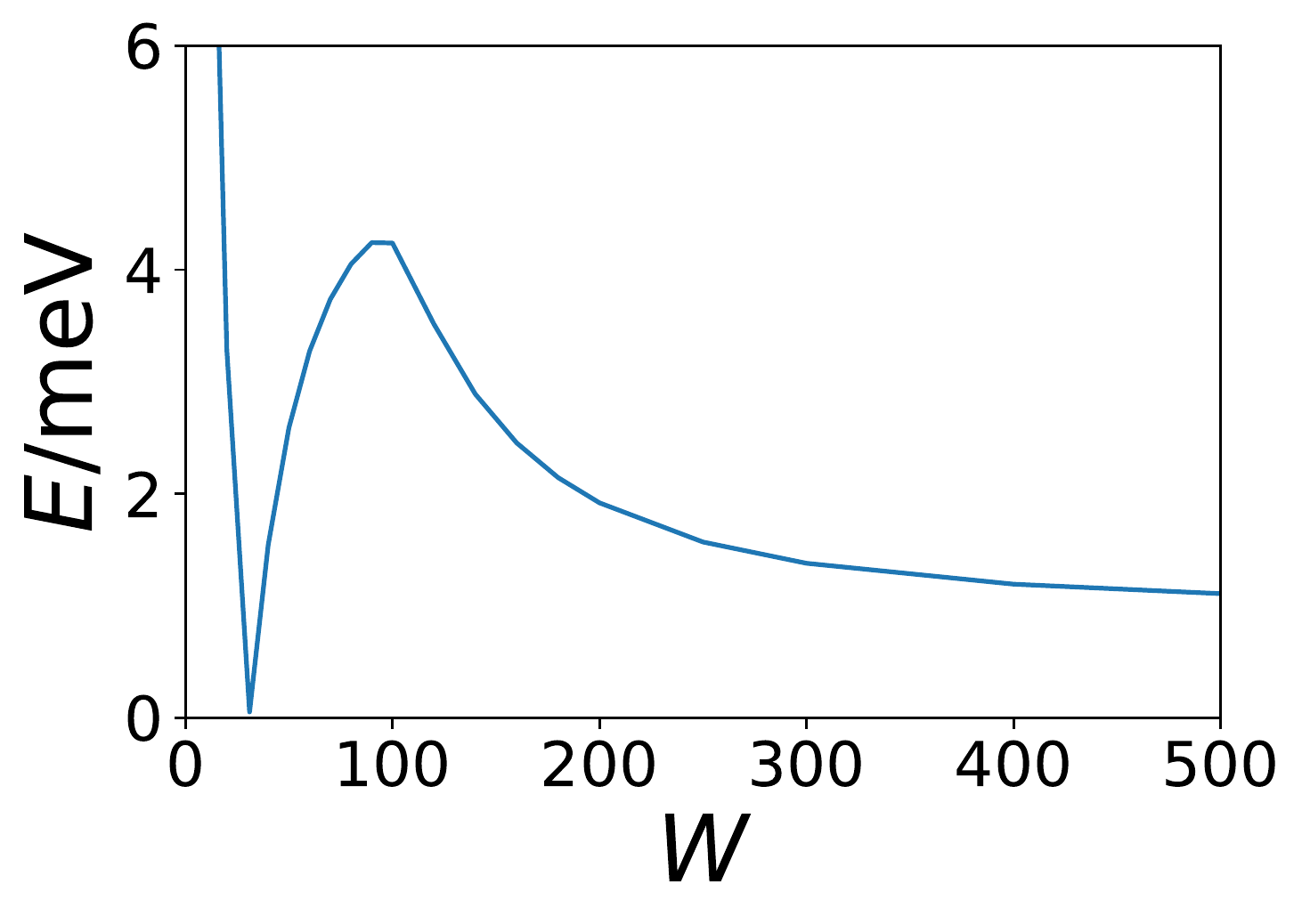}
		\caption{} \label{gap_width_v}
	\end{subfigure}
	\caption{(a) Phase diagram in $\lambda$-$\mu$ space for a quasi-1D chain with $W=300, m_0=\SI{6}{\milli\electronvolt}, m_1=\SI{0.1}{\electronvolt}, v=\SI{0.2}{\electronvolt}$, $\lambda'=0$, $\Delta_b=\SI{1}{\milli\electronvolt}, \Delta_t=0$;
	(b) Phase diagram in $\lambda'$-$\mu$ space with $\lambda=\SI{12}{meV}$ and other parameters the same as in (a);
	(c) Relation between gap and width, with $\mu=\SI{5}{\milli\electronvolt},\lambda=\SI{12}{meV}$ and other parameters the same as in (a).} \label{lower_band}
\end{figure*}

Fig. \ref{lower_band} shows results from calculation with lower-band parameters. The phase diagrams show more narrow alternating patterns, but near $\mu=0$ large topological regions still exist. Large $v, m_1$ and small $\Delta$ ensures a large gap protecting MZMs. Therefore, as long as $v,m_1\gg m_0,\Delta$, our results in the main text always hold true, and the proposed experimental scheme is always feasible.

\section{Spatial Distribution}
\begin{figure}[H]
\centering
\includegraphics[width=\linewidth]{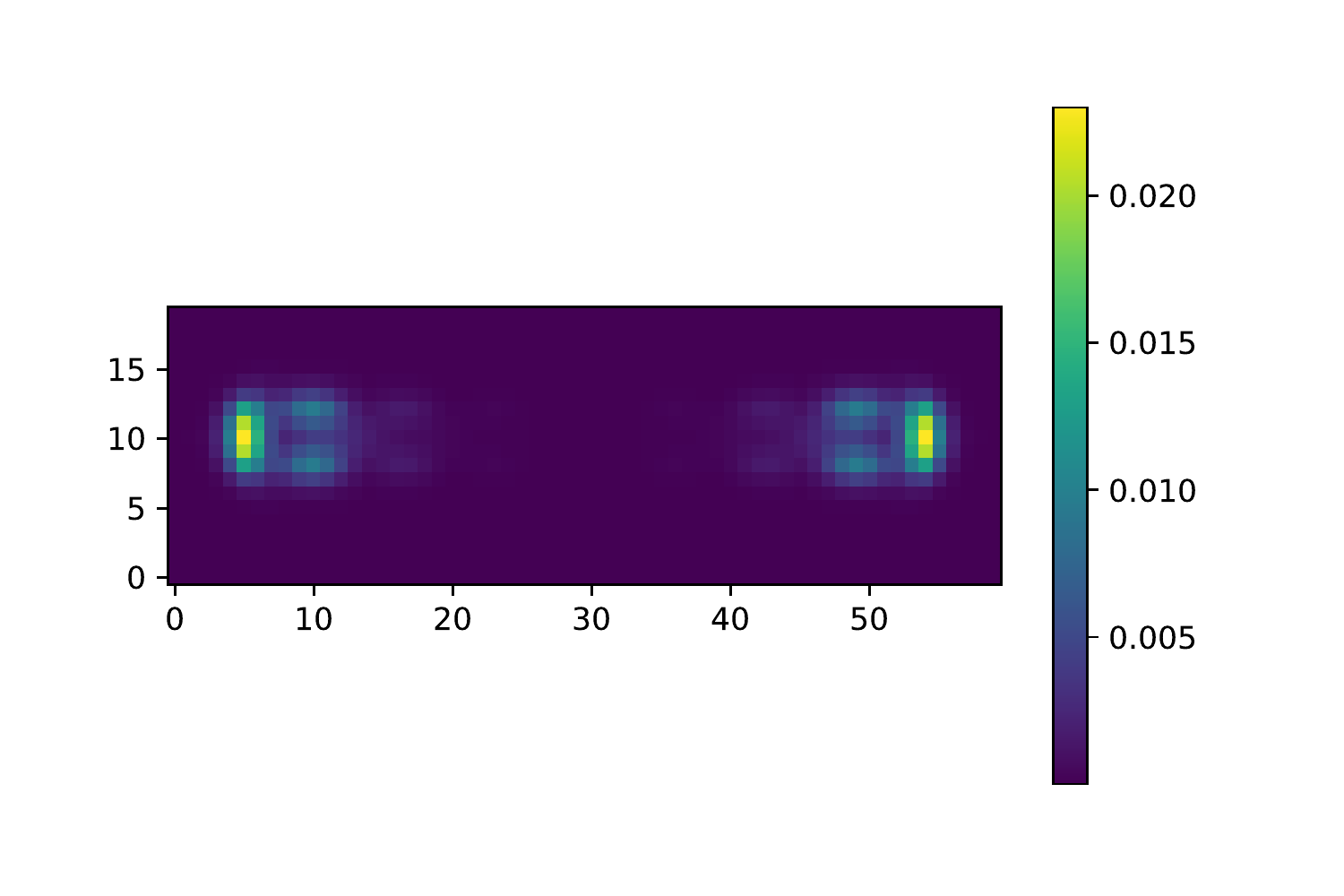}
\caption{Spatial distribution of the lowest energy eigenstate, where a $50\times5$ quasi-1D region is embedded in a $60\times20$ system, and $m_0=1, m_1=2, v=2$, $\lambda=2$, $\mu=1, \Delta_t=1, \Delta_b=0$. Inside and outside the quasi-1D region, $\lambda'=0$ and 4 respectively.} \label{wavefunction}
\end{figure}

Now we construct a finite system and calculate the spatial distribution of MZM. A topological quasi-1D region is embedded in the system, surrounded by regions with large $\lambda'$. For numerical convenience, here we choose some fictitious parameters with small $v$ and $m_1$, instead of more realistic ones. The result is shown in Fig. \ref{wavefunction}. The numerical calculation was performed using the Kwant code \cite{groth2014kwant}.

\bibliography{bib_majo}

\end{document}